\documentclass[a4paper,10pt]{article}
\usepackage{graphicx}
\usepackage{comment}
\usepackage{amssymb}
\usepackage{amsmath}
\usepackage{nicefrac}
\usepackage{graphicx}
\usepackage{dcolumn}
\usepackage{bm}
\usepackage{comment}
\usepackage{multirow}
\usepackage{cite}
\usepackage{url}

\newcommand{\bin}[2]{\left(\begin{array}{c}\!#1\!\\\!#2\!\end{array}\right)}
\newcommand{\threejm}[6]{\left(\begin{array}{ccc}#1 & #2 & #3 \\ #4 & #5 & #6 \end{array}\right)}

\def\anglefig{0}
\def\scalefig{0.90}

\begin{document}

\huge

\begin{center}
Complex pseudo-partition functions in the Configurationally-Resolved Super-Transition-Array approach for radiative opacity
\end{center}

\vspace{0.3cm}

\large

\begin{center}
Jean-Christophe Pain$^{a,b,}$\footnote{jean-christophe.pain@cea.fr}
\end{center}

\normalsize

\begin{center}
\it $^a$CEA, DAM, DIF, F-91297 Arpajon, France\\
\it $^b$Universit\'e Paris-Saclay, CEA, Laboratoire Mati\`ere en Conditions Extr\^emes,\\
\it 91680 Bruy\`eres-le-Ch\^atel, France
\end{center}

\vspace{0.3cm}

\begin{abstract}
A few years ago, Kurzweil and Hazak developed the Configurationally Resolved Super-Transition-Arrays (CRSTA) method for the computation of hot-plasma radiative opacity. Their approach, based on a temporal integration, is an important refinement of the standard Super-Transition-Arrays (STA) approach, which enables one to recover the underlying structure of the STAs, made of unresolved transition arrays. The CRSTA formalism relies on the use of complex pseudo partition functions, depending on the considered one-electron jump. In this article, we find that, despite the imaginary part, the doubly-recursive relation which was introduced in the original STA method to avoid problems due to alternating-sign terms in partition functions, is still applicable, robust, efficient, and exempt of numerical instabilities. This was rather unexpected, in particular because of the occurrence of trigonometric functions, or Chebyshev polynomials, which can be either positive or negative. We also show that, in the complex case, the recursion relation can be presented in a form where the vector of real and imaginary parts at a given iteration is therefore obtained by a sum of the rotated previous ones. 
\end{abstract}

\section{Introduction}\label{sec1}

The Configurationally Resolved Super-Transition-Arrays (CRSTA) method for the computation of radiative opacity was proposed by Kurzweil and Hazak \cite{Hazak2012,Kurzweil2013}. Such an approach is probably the most important theoretical improvement since the pioneering work of Bar-Shalom \emph{et al.} introducing the Super Transition Arrays (STA) statistical model \cite{Barshalom1989}, which has proven successful for the interpretation of spectroscopy experiments \cite{Kurzweil2022} and was implemented in several efficient opacity codes \cite{Krief2015,Krief2018,Aberg2021,Pain2022,Gill2023b}. However, in some cases, the STA spectrum is not sufficiently resolved, which is important for improving the interpretation of transmission measurements or for obtaining accurate Rosseland mean opacities, key ingredients of radiative-transfer modeling.

A CRSTA represents the exact sum of the spectra of all Unresolved Transition Arrays (UTAs) \cite{Bauche1988} constituting the STA and sharing the same set of self-consistent-field one-electron energies and wavefunctions. The main advantage of the method is that it enables one to avoid the STA Gaussianity assumption The radiation intensity, average energy and variance of the original STA are recovered by truncating a cumulant expansion at the third term \cite{Hazak2012}. In other words, the STA is the coarse-grained Gaussian approximation of the spectrum of the corresponding CRSTA. Within the CRSTA formalism, the photo-absorption cross-section reads (without the stimulated-emission correction):
\begin{eqnarray}\label{basic}
    \sigma(\omega)&=&\frac{4\pi^2\alpha}{3}\hbar\omega\frac{1}{\mathcal{Z}(g,\beta,0)}\sum_{\Xi,a,b}\langle a||r||b\rangle^2g_ag_b\nonumber\\
    & &\times\Re\left\{\int_0^{\infty}e^{i\left(D_0^{ab}-D_a^{ab}\right)\tau/\hbar}\tilde{X}_a^{ab}(\beta,\tau)\mathcal{Z}_{\Xi,Q-1}^{ab}\left(g-\delta_a-\delta_b,\beta,\tau\right)e^{-i\omega\tau}d\tau\right\}
\end{eqnarray}
with
\begin{equation}
    D_0^{ab}=I_b-I_a
\end{equation}
where
\begin{equation}
	I_s=\int_0^{\infty}\left(\epsilon_s-eV(r)-\frac{Ze^2}{r}\right)\left[P_{n_s,\ell_s}(r)\right]^2dr
\end{equation}
and $P_{n_s,\ell_s}$ is the radial part of the wavefunction of the one-electron state $s$ with energy $\epsilon_s$. The degeneracy of subshell $j$ is denoted $g_j$. $g$ denotes the vector of degeneracies $\left\{g_1, g_2, \cdots, g_N\right\}$. $\alpha$ represents the fine-structure constant and $\langle a||r||b\rangle$ the dipole matrix element. $\Re(z)$ is the real part of $z$. In addition,
\begin{equation}\label{dmat}
    D_s^{ab}=V_{s,b}-V_{s,a},
\end{equation}
where for non-equivalent electrons ($r\ne s$, see Ref. \cite{Cowan1981}, p. 165) :
\begin{equation}\label{noneq}
    V_{s,r}=F^{(0)}\left(n_s\ell_s,n_s\ell_r\right)-\frac{1}{2}\sum_{k\ge 0}\threejm{\ell_r}{k}{\ell_s}{0}{0}{0}^2G^{(k)}(n_r,\ell_r,n_s\ell_s)
\end{equation}
and for equivalent electrons 
\begin{equation}\label{eq}
    V_{r,r}=F^{(0)}\left(n_r\ell_r,n_r\ell_r\right)-\frac{(2\ell_r+1)}{(4\ell_r+1)}\sum_{k>0}\threejm{\ell_r}{k}{\ell_r}{0}{0}{0}^2F^{(k)}(n_r,\ell_r,n_r\ell_r).
\end{equation}
Equations (\ref{noneq}) and (\ref{eq}) can be gathered into (see \ref{appA}):
\begin{equation}
    \tilde{V}_{s,r}=F^{(0)}\left(n_s\ell_s,n_r\ell_r\right)-(1-\delta_{s,r})\frac{\delta_{\ell_r,\ell_s}}{g_s}G^{(0)}(n_s\ell_s,n_r\ell_r)-\frac{g_s}{2(g_s-\delta_{s,r})}\sum_{k>0}\threejm{\ell_s}{k}{\ell_r}{0}{0}{0}^2G^{(k)}(n_s\ell_s,n_r,\ell_r).
\end{equation}
where $F^{(k)}$ and $G^{(k)}$ are the direct and exchange Slater integrals. An important specificity of the CRSTA method is that it involves complex pseudo-partition functions (``complex'' in the sense that they have non-zero real and imaginary parts) and ``pseudo'' because they depend on a specific mono-electron jump (transition) $a\rightarrow b$ (simply denoted $ab$). Furthermore, they have to be integrated (after multiplication by other quantities) over $\tau$ (see Eq. (\ref{basic})). One has
\begin{equation}
    \mathcal{Z}(g,\beta,\tau)=\sum_{ab}\mathcal{Z}^{ab}(g,\beta,\tau)
\end{equation}
with (the sum runs over all superconfigurations $\Xi$):
\begin{equation}
    \mathcal{Z}^{ab}(g,\beta,\tau)=\sum_{\Xi}\mathcal{Z}_{\Xi}^{ab}(g,\beta,\tau)
\end{equation}
and (the sum runs over all configurations $C$ belonging to the superconfiguration $\Xi$):
\begin{equation}
    \mathcal{Z}_{\Xi}^{ab}\left(g,\beta,\tau\right)=\sum_{C\in\Xi}\prod_{s\in C}\bin{g_s}{q_s}\left[\tilde{X}_s^{ab}(\beta,\tau)\right]^{q_s}.
\end{equation}
We also introduce
\begin{equation}
    \mathcal{Z}_{\Xi,Q}^{ab}\left(g,\beta,\tau\right)=\sum_{\substack{C\in\Xi\\\sum_{s\in C}q_s=Q}}
    \prod_{s\in C}\bin{g_s}{q_s}\left[\tilde{X}_s^{ab}(\beta,\tau)\right]^{q_s},
\end{equation}
with
\begin{equation}\label{eqq35}
    \tilde{X}_s^{ab}(\beta,\tau)=\exp\left[-\beta(\epsilon_s-\mu)+i\,D_s^{ab}\frac{\tau}{\hbar}\right].
\end{equation}
We can write
\begin{equation}
 \tilde{X}_s^{ab}(\beta,\tau)=X_s\exp\left[i\,D_s^{ab}\frac{\tau}{\hbar}\right],
\end{equation}
where $X_s=\exp\left[-\beta(\epsilon_s-\mu)\right]$ is the usual factor of the STA theory, with the notation $\beta=1/(k_BT)$, $k_B$ being the Boltzmann constant. $\mu$ is the chemical potential.
 
It is possible to account for the homogeneous broadening and UTA widths in the framework of the CRSTA method (see section \ref{sec3}), although the corresponding terms are not included in Eq. (\ref{eqq35}).

In the STA approach, the partition function of a supershell with $N$ subshells and $Q$ electrons is defined as
\begin{equation}
    \mathcal{U}_Q(g)=\sum_{\substack{\{p_s\}\\\sum_{s=1}^Np_s=Q}}\prod_{s=1}^N\bin{g_s}{p_s}X_s^{p_s}.
\end{equation}
We use the notation
\begin{equation}
    \sum_{\substack{\{p_s\}\\\sum_{s=1}^Np_s=Q}}\cdots=\underbrace{\sum_{q_1=0}^{g_1}\sum_{q_2=0}^{g_2}\cdots \sum_{q_N=0}^{g_N}}_{\sum_{s=1}^Np_s=Q}\cdots,
\end{equation}
and throughout the paper, when all the quantities in a given relation involve the same set of degeneracies, we omit it as an argument (i.e., for instance, we replace $\mathcal{U}_Q(g)$ by $\mathcal{U}_Q$). The total degeneracy of a supershell will be denoted
\begin{equation}
G=\sum_{i=1}^Ng_i.
\end{equation}

\begin{table}
\begin{center}
\begin{tabular}{ccc}\hline\hline
 Subshell & $\epsilon_{n\ell}$ (eV) & $D_{n\ell}^{\mathrm{3d,4f}}$ (eV)\\\hline\hline
   4p & -1337.9851 & -17.18491\\
   4d & -1107.9064 & -17.41130\\
   4f & -862.92791 & -15.71795\\
   5s & -735.12539 & -5.27529\\
   5p & -660.63832 & -4.95181\\
   5d & -566.64961 & -4.91479\\\hline\hline
\end{tabular}
\end{center}
\caption{One-electron subshell energies and $D$ matrix (see Eq. (\ref{dmat})) for a gold plasma at $T$=100 eV and $\rho$=0.01 g/cm$^3$. The chemical potential is $\mu$=-895.58476 eV. Such quantities were obtained from a self-consistent average-atom calculation \cite{Pain2007}.}\label{tab1}
\end{table}

\begin{figure}[!ht]
\vspace{1cm}
 \begin{minipage}[c]{0.48\textwidth}
   \centering
   \includegraphics[width=\textwidth,angle=\anglefig,scale=\scalefig]{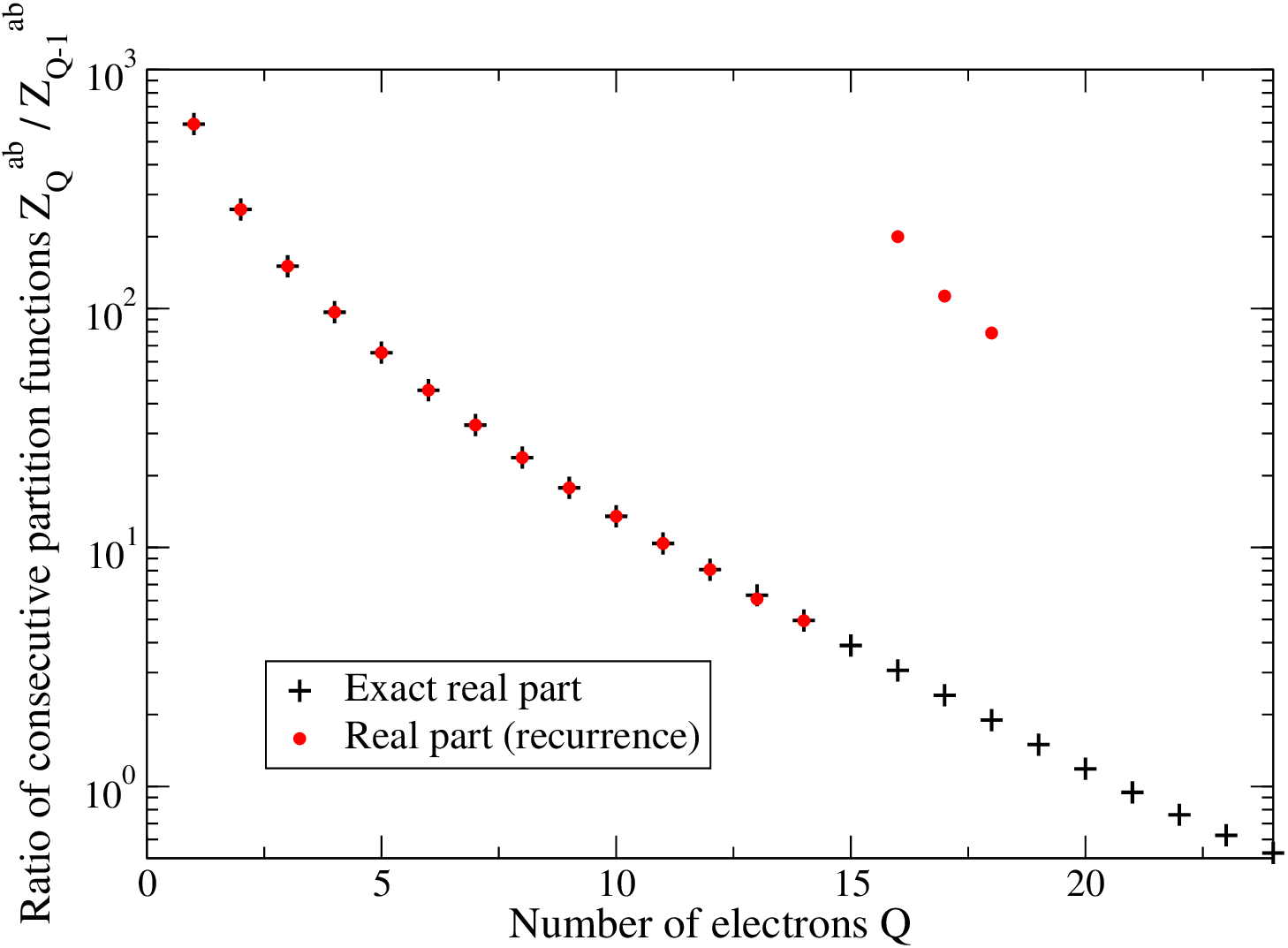}
   \caption{Real part of the ratio of consecutive partition functions $\mathcal{Z}_Q^{ab}/\mathcal{Z}_{Q-1}^{ab}$ for $\tau=10$ in the case of a copper plasma at $T$=100 eV and mass density $\rho$=0.01 g/cm$^3$, computed using the Bar-Shalom relation (\ref{red}) \cite{Barshalom1989}. The considered supershell is (4p4d4f5s5p5d). The chemical potential is $\mu$=-895.58476 eV. The one-electron energies and $D$-matrix elements are provided in Table \ref{tab1}. Owing to the particle-hole symmetry, we focus on the numbers of electrons lower or equal to $G/2$=24.}\label{fig3_bs_real}
 \end{minipage}\hfill
 \begin{minipage}[c]{0.48\textwidth}
   \centering
   \includegraphics[width=\textwidth,angle=\anglefig,scale=\scalefig]{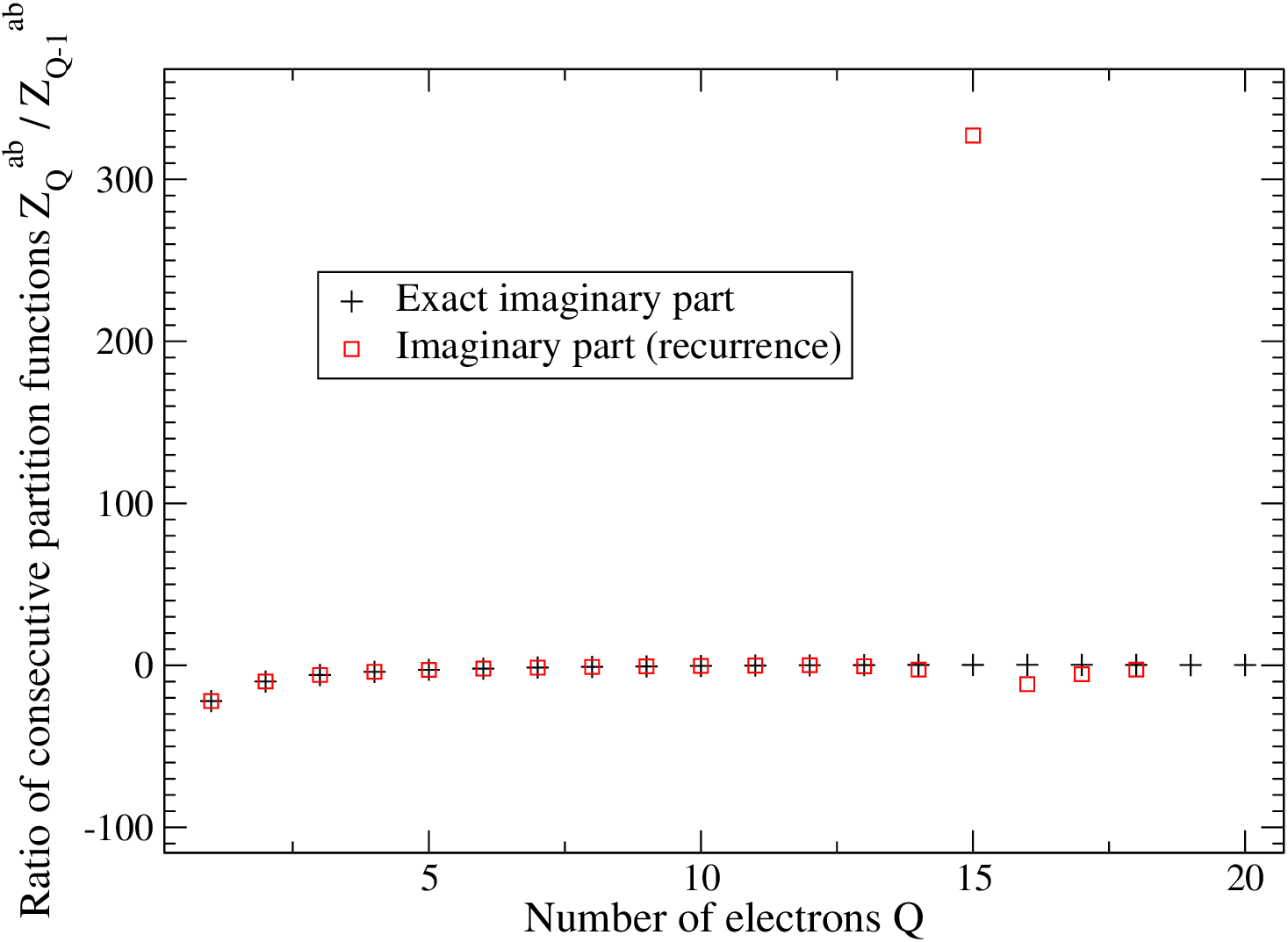}
   \caption{Imaginary part of the ratio of consecutive partition functions $\mathcal{Z}_Q^{ab}/\mathcal{Z}_{Q-1}^{ab}$ for $\tau=10$ in same conditions as Fig. \ref{fig3_bs_real}.}\label{fig3_bs_imag}
 \end{minipage}
\end{figure}

The calculation of the complex pseudo partition functions involved in the CRSTA method is investigated in the next section. The application of the usual Bar-Shalom relation is presented in subsection \ref{subsec21}, the new variant of the doubly-recursive relation is explained in subsection \ref{subsec22}, and the generalization to a recently-published asymptotic approximation is being considered in subsection \ref{subsec23}. The specific nature of application of the present formalism to partially resolved super transition arrays is discussed in section \ref{sec3}.

\section{Calculation of the complex pseudo partition functions}\label{sec2}

\subsection{The Bar-Shalom relation}\label{subsec21}

Let us set
\begin{equation}
    \mathcal{K}_Q(a,b)=\Re\left\{\int_0^{\infty}e^{i\left(D_0^{ab}-D_a^{ab}\right)\tau/\hbar}\tilde{X}_a^{ab}(\beta,\tau)\mathcal{Z}_{\Xi,Q-1}^{ab}\left(g-\delta_a-\delta_b,\beta,\tau\right)e^{-i\hbar\omega\frac{\tau}{\hbar}}d\tau\right\}.
\end{equation}
Equation (\ref{basic}) becomes
\begin{equation}
    \sigma(\omega)=\frac{4\pi^2\alpha}{3}\hbar\omega\frac{1}{\mathcal{Z}(g,\beta,0)}\sum_{\Xi,a,b}\langle a||r||b\rangle^2g_ag_b\times\mathcal{K}_Q(a,b).
\end{equation}

Using the standard Bar-Shalom relation (see \ref{appB}), one finds
\begin{equation}\label{red}
    \mathcal{Z}_{Q-1}^{ab}(g^{ab},\beta,\tau)=\frac{1}{Q-1}\sum_{n=0}^{Q-2}\chi_{Q-1-n}^{ab}\mathcal{Z}_n^{ab}(g^{ab},\beta,\tau)=\frac{1}{Q-1}\sum_{n=1}^{Q-1}\chi_{n}^{ab}\mathcal{Z}_{Q-1-n}^{ab}(g^{ab},\beta,\tau)
\end{equation}
with
\begin{equation}\label{chi}
    \chi_n^{ab}=-\sum_{j=1}^Ng_j\left[-\tilde{X}_j^{ab}(\beta,\tau)\right]^n.
\end{equation}
In Eq. (\ref{red}), we have introduced the notation $g^{x_1x_2x_3\cdots}=g-\delta_{x_1}-\delta_{x_2}-\delta_{x_3}-\cdots$. One has also
\begin{equation}
    \mathcal{Z}_{Q-1}^{ab}(g^{ab},\beta,\tau)=\sum_{n=0}^{Q-1}\left[-\tilde{X}_b^{ab}(\beta,\tau)\right]^{Q-1-n}\mathcal{Z}_n^{ab}(g^a,\beta,\tau),
\end{equation}
which becomes, using
\begin{equation}
    \mathcal{Z}_{n}^{a}(g^{ab},\beta,\tau)=\sum_{k=0}^{n}\left[-\tilde{X}_a^{ab}(\beta,\tau)\right]^{n-k}\mathcal{Z}_k^{ab}(g,\beta,\tau),
\end{equation}
the relation
\begin{equation}
    \mathcal{Z}_{Q-1}^{ab}(g^{ab},\beta,\tau)=\sum_{n=0}^{Q-1}\sum_{k=0}^{n}(-1)^{Q-1-k}\left[\tilde{X}_a^{ab}(\beta,\tau)\right]^{n-k}\left[\tilde{X}_b^{ab}(\beta,\tau)\right]^{Q-1-n}\mathcal{Z}_k^{ab}(g,\beta,\tau),
\end{equation}
which is a particular case of
\begin{equation}
    \mathcal{Z}_{Q}^{ab}(g^{a_1a_2\cdots a_p},\beta,\tau)=\sum_{n_1=0}^{Q-1}\left[-\tilde{X}_{a_1}^{ab}(\beta,\tau)\right]^{Q-n_1}\sum_{n_2=0}^{n_1}\left[-\tilde{X}_{a_2}^{ab}(\beta,\tau)\right]^{n_1-n_2}\cdots\sum_{n_p=0}^{n_{p-1}}\left[-\tilde{X}_{a_p}^{ab}(\beta,\tau)\right]^{n_{p-1}-n_p}\mathcal{Z}^{ab}_{n_p}(g,\beta,\tau).
\end{equation}
The factor $\mathcal{K}$ entering the expression of the absorption coefficient is therefore
\begin{equation}
    \mathcal{K}_Q(a,b)=\sum_{n=0}^{Q-1}\sum_{k=0}^{n}(-1)^{Q-1-k}\Re\left\{\int_0^{\infty}e^{i\left(D_0^{ab}-D_a^{ab}\right)\tau/\hbar}\left[X_a^{ab}(\beta,\tau)\right]^{n-k}\left[X_b^{ab}(\beta,\tau)\right]^{Q-1-n}\mathcal{Z}_k^{ab}(g,\beta,\tau)e^{-i\hbar\omega\tau/\hbar}d\tau\right\}.
\end{equation}

Because to the alternate signs in Eqs. (\ref{red}) (due to Eq. (\ref{chi})), numerical errors may happen for the calculation of the complex pseudo partition function $\mathcal{Z}_Q$, as in the usual (STA) real case $\mathcal{U}_Q$, as can bee seen in Figs. \ref{fig3_bs_real} and \ref{fig3_bs_imag} for $\tau=10$. Throughout the paper we consider the case of a copper plasma at $T$=100 eV and mass density $\rho$=0.01 g/cm$^3$. The considered supershell is (4p4d4f5s5p5d) with a total degeneracy $G$=48. The one-electron energies and $D$-matrix elements are provided in Table \ref{tab1}, together with the chemical potential. Since all the partition-function algorithms can be optimized with the particle-hole symmetry, an algorithm is considered numerically stable and robust if it works properly for a number of electrons from one to $G/2$=24. We can see that both the real and imaginary parts are unstable in that case.

\begin{figure}[!ht]
 \begin{minipage}[c]{0.48\textwidth}
   \centering
   \includegraphics[width=\textwidth,angle=\anglefig,scale=\scalefig]{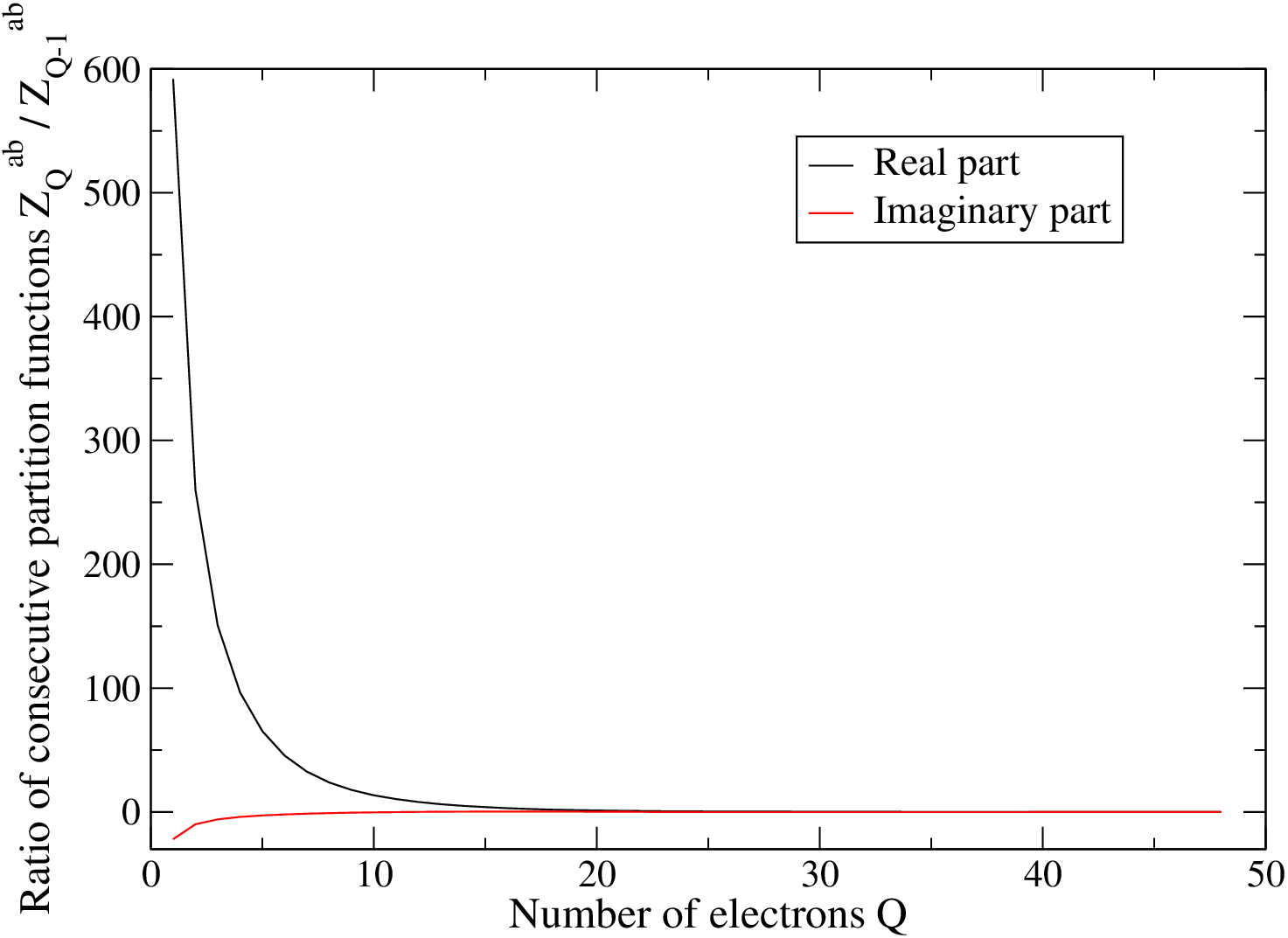}
   \caption{Real part of the ratio of consecutive partition functions $\mathcal{Z}_Q^{ab}/\mathcal{Z}_{Q-1}^{ab}$ for $\tau=10$ in the case of a copper plasma at $T$=100 eV and mass density $\rho$=0.01 g/cm$^3$, computed using the doubly-recursive relation (\ref{double}) \cite{Gilleron2004}. The considered supershell is (4p4d4f5s5p5d). The chemical potential is $\mu$=-895.58476 eV. The one-electron energies and $D$-matrix elements are provided in Table \ref{tab1}.}\label{fig1_full}
 \end{minipage}\hfill
 \begin{minipage}[c]{0.48\textwidth}
   \centering
   \includegraphics[width=\textwidth,angle=\anglefig,scale=\scalefig]{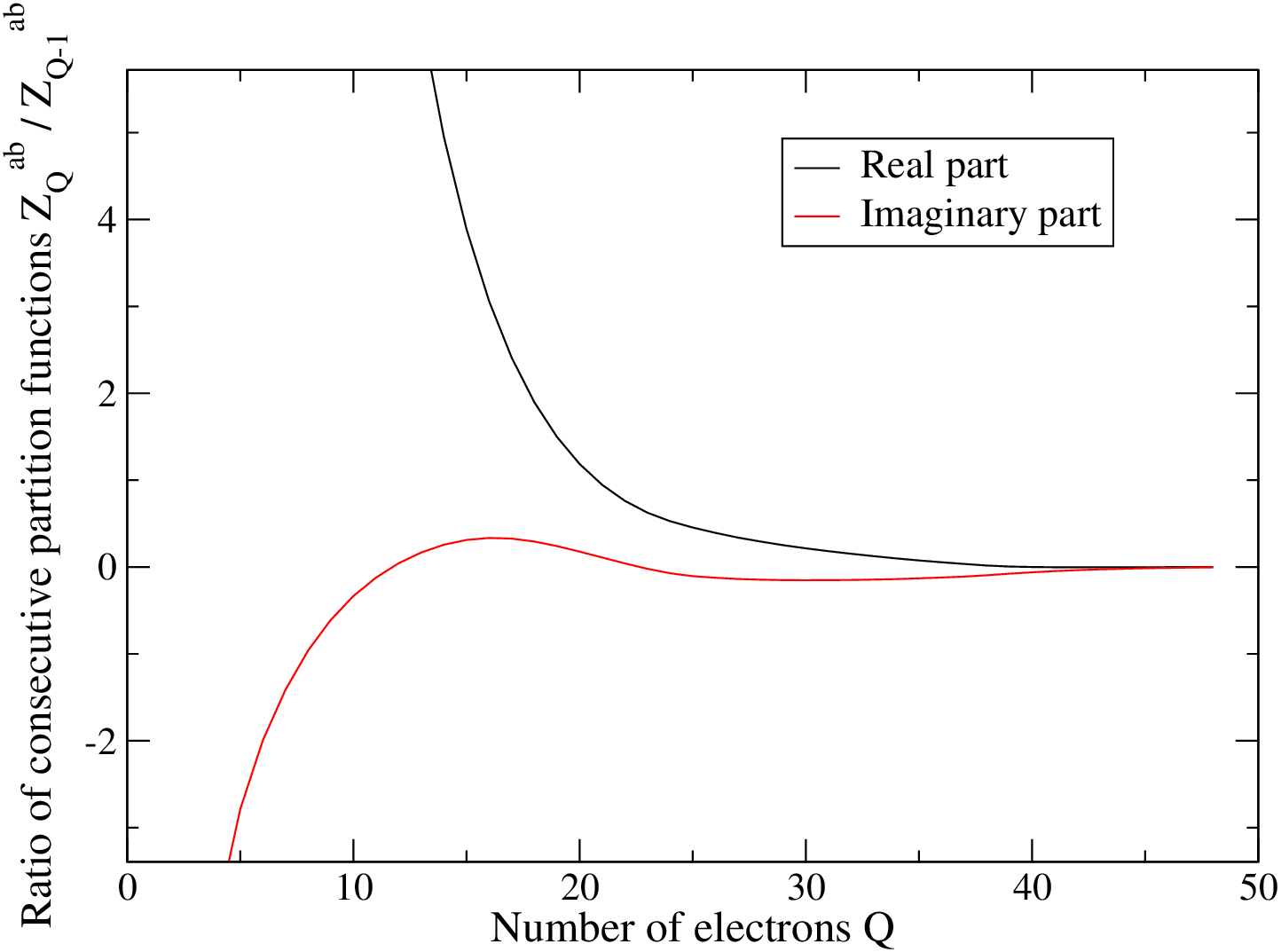}
   \caption{Zoom of Fig. \ref{fig1_full} on a reduced range of the number of electrons.}\label{fig1_bis}
 \end{minipage}
\end{figure}

\vspace{1cm}

\begin{figure}[!ht]
 \begin{minipage}[c]{0.48\textwidth}
   \centering
   \includegraphics[width=\textwidth,angle=\anglefig,scale=\scalefig]{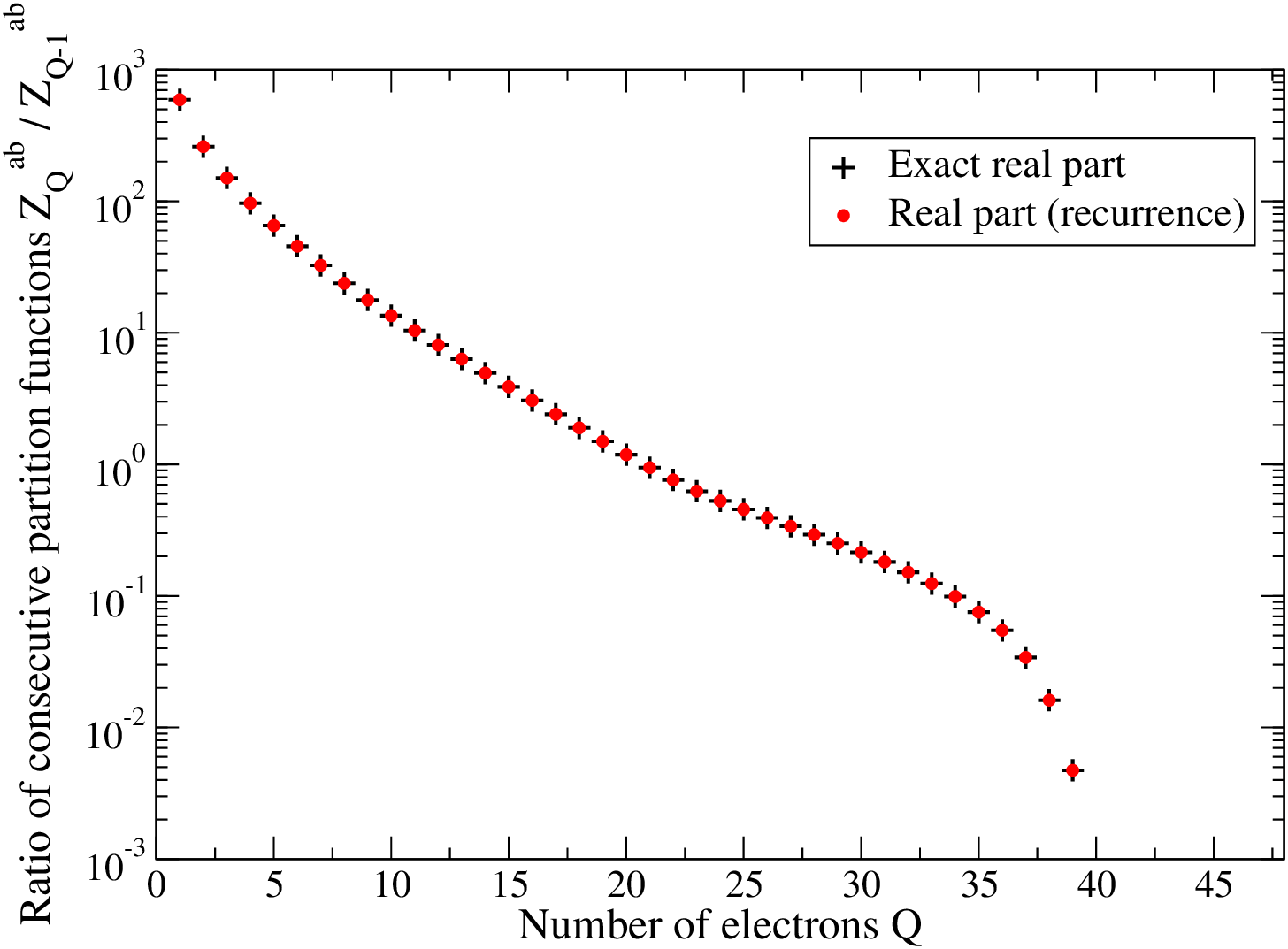}
   \caption{Real part of the ratio displayed in Figs. \ref{fig1_full} and \ref{fig1_bis} but on a logarithmic scale (most common representation). After 40 electrons, the values become negative. Results are superimposed with the exact ones.}\label{fig2_real}
 \end{minipage}\hfill
 \begin{minipage}[c]{0.48\textwidth}
   \centering
   \includegraphics[width=\textwidth,angle=\anglefig,scale=\scalefig]{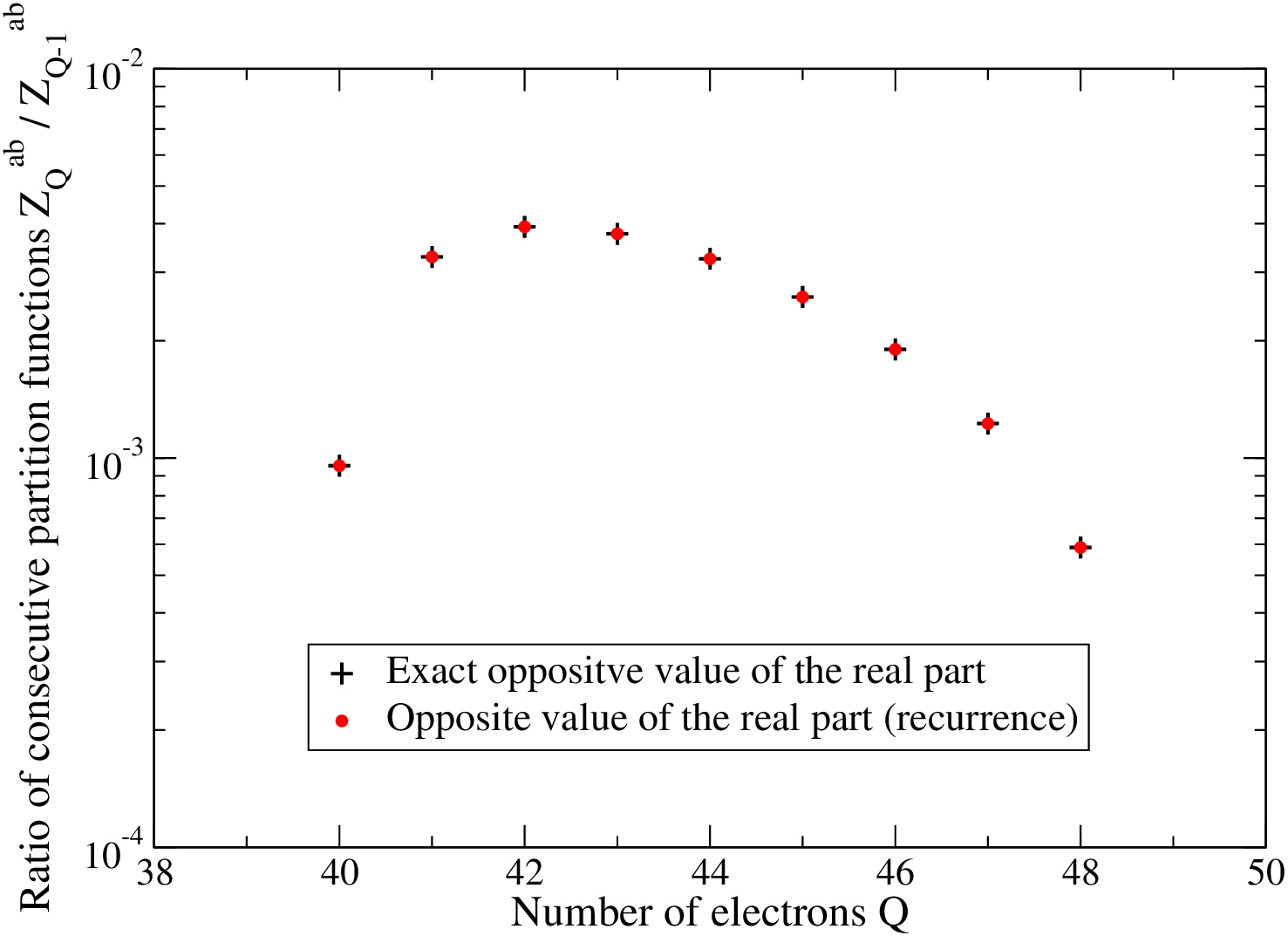}
   \caption{Absolute value of the real part of the ratio displayed in Figs. \ref{fig1_full} and \ref{fig1_bis}, for a number of electrons greater or equal to 40. Results are superimposed with the exact ones.}\label{fig2_real_bis}
 \end{minipage}
\end{figure}

\vspace{1cm}

\begin{figure}
\begin{center}
\includegraphics[scale=0.35]{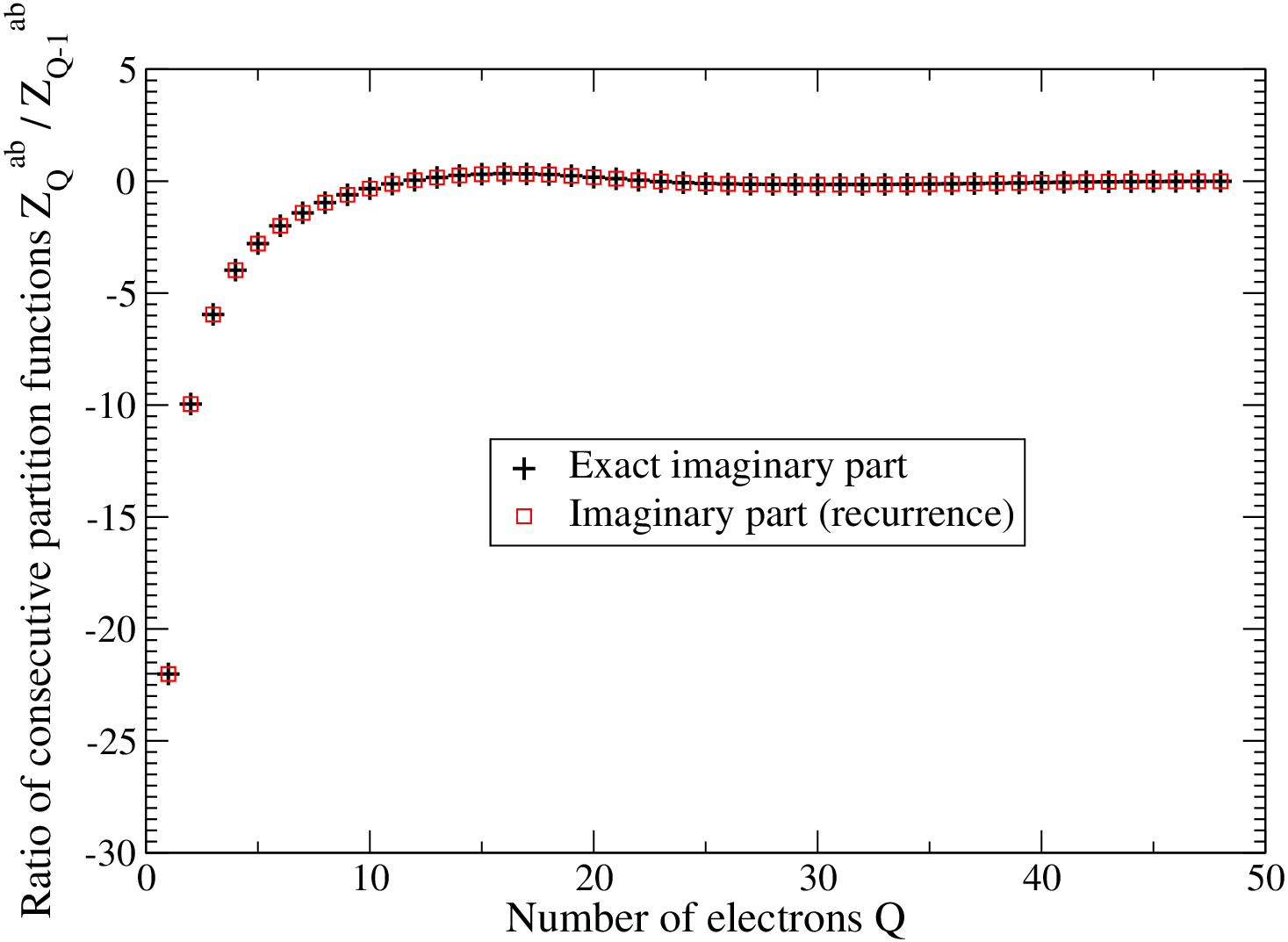}
\caption{Imaginary part of the ratio of consecutive partition functions $\mathcal{Z}_Q^{ab}/\mathcal{Z}_{Q-1}^{ab}$ for $\tau=10$ in the case of a copper plasma at $T$=100 eV and mass density $\rho$=0.01 g/cm$^3$, computed using the doubly-recursive relation (\ref{double}) \cite{Gilleron2004}. The considered supershell is (4p4d4f5s5p5d). The chemical potential is $\mu$=-895.58476 eV. The one-electron energies and $D$-matrix elements are provided in Table \ref{tab1}. Results are superimposed with the exact ones.}\label{fig2_imag}
\end{center}
\vspace{1cm}
\end{figure}

\subsection{The doubly-recursive relation}\label{subsec22}

In order to remedy the fact that the recursion relations can be unstable in the STA theory, we proposed in 2004 a relation made of a recursion over the both numbers of electrons and subshells \cite{Gilleron2004,Wilson2007,Pain2020,Pain2021}. It consists in a factorization of the contribution of a specific subshell (in the present work we will focus on the last one for simplicity):
\begin{equation}\label{double}
	\mathcal{U}_{Q,N}=\sum_{p=0}^{\min(Q,g_N)}X_N^pe^{ip\alpha_N}\mathcal{U}_{Q-p,N-1}.
\end{equation}
For simplicity, let us change the notation $\mathcal{Z}_{Q-1}^{ab}(g,\beta,\tau)$ into $\mathcal{Z}_{Q,N}$ (we omit the superscripts $a$ and $b$ as well as the dependencies with respect to $\beta$ and $\tau$, which are all always present), and introduce explicitly the dependency with respect to the number of subshells $N$. Within the CRSTA approach, the above relation (\ref{double}) becomes
\begin{equation}
    \mathcal{Z}_{Q,N}=\sum_{p=0}^{\min(Q,g_N)}X_N^pe^{ip\alpha_N}\mathcal{Z}_{Q-p,N-1}.
\end{equation}

When dealing with complex numbers, some precautions have to be taken. Complex multi-precision multiplication is performed using the identity
\begin{equation}
    (a_1+i\,a_2)(b_1+i\,b_2)=(a_1b_1-a_2b_2)+i\left[(a_1+a_2)(b_1+b_2)-a_1b_1-a_2b_2\right]
\end{equation}
This formula can be implemented using only three multiprecision multiplications, whereas the straightforward formula requires four. Complex division is performed using the identity
\begin{equation}
    \frac{a_1+i\,a_2}{b_1+i\,b_2}=\frac{(a_1+i\,a_2)(b_1-i\,b_2)}{b_1^2+b_2^2}
\end{equation}
where the complex product in the numerator is evaluated as above. Since division is significantly more expensive than multiplication, the two real divisions ordinarily required in this formula are replaced with a reciprocal computation of $b_1^2+b_2^2$ followed by two multiplications. The advanced routines for complex multiplication and division utilize these same formulas, but they call the advanced routines for real multiplication and division \cite{Bailey1993}. Van Meter and Itoh have performed a detailed analysis of the impact on quantum modular exponentiation of quantum-computer architectures and possible concurrent-gate execution. They found that exponentiation of a $n$-bit number can be executed a few hundreds times faster than optimized versions of the basic algorithms. On a neighbor-only architecture, time is $O(n)$, whereas non-neighbor architectures can reach $O(\log n)$ \cite{VanMeter2005}.

Let us introduce the real and imaginary parts $\mathcal{A}_{Q,N}$ and $\mathcal{B}_{Q,N}$ of the complex pseudo partition function $\mathcal{Z}_{Q,N}$:
\begin{equation}
    \mathcal{Z}_{Q,N}=\mathcal{A}_{Q,N}+i\,\mathcal{B}_{Q,N}.
\end{equation}
In the present case, one has
\begin{eqnarray}
    \mathcal{Z}_{Q,N}&=&\sum_{p=0}^{\min(Q,g_N)}X_N^pe^{ip\alpha_N}\mathcal{Z}_{Q-p,N-1}=\sum_{p=0}^{\min(Q,g_N)}X_N^p\left(\mathcal{A}_{Q-p,N-1}+i\,\mathcal{B}_{Q-p,N-1}\right)\nonumber\\
    &=&\sum_{p=0}^{\min(Q,g_N)}X_N^p\left[\cos(p\alpha_N)+i\,\sin(p\alpha_N)\right]\left(\mathcal{A}_{Q-p,N-1}+i\,\mathcal{B}_{Q-p,N-1}\right)
\end{eqnarray}
or equivalently
\begin{eqnarray}
    \mathcal{Z}_{Q,N}&=&\sum_{p=0}^{\min(Q,g_N)}X_N^p\left(\cos(p\alpha_N)\mathcal{A}_{Q-p,N-1}-\sin(p\alpha_N)\mathcal{B}_{Q-p,N-1}\right)\nonumber\\
    & &+i\,\sum_{p=0}^{\min(Q,g_N)}X_N^p\left[\cos(p\alpha_N)\mathcal{B}_{Q-p,N-1}+\sin(p\alpha_N)\mathcal{A}_{Q-p,N-1}\right].
\end{eqnarray}

Identifying the real and imaginary parts in the left- and right-hand sides of the preceding equation, we finally get the system of coupled equations
\begin{equation}\label{rea}
    \mathcal{B}_{Q,N}=\sum_{p=0}^{\min(Q,g_N)}X_N^p\left[\cos(p\alpha_N)\mathcal{A}_{Q-p,N-1}-\sin(p\alpha_N)\mathcal{B}_{Q-p,N-1}\right]
\end{equation}
and
\begin{equation}\label{ima}
    \mathcal{A}_{Q,N}=\sum_{p=0}^{\min(Q,g_N)}X_N^p\left[\cos(p\alpha_N)\mathcal{B}_{Q-p,N-1}+\sin(p\alpha_N)\mathcal{A}_{Q-p,N-1}\right].
\end{equation}
Figure \ref{fig1_full} illustrates the real part of the ratio between successive partition functions, $\mathcal{Z}_Q^{ab}/\mathcal{Z}_{Q-1}^{ab}$, computed for $\tau = 10$ in a copper plasma at a temperature of 100 eV and a mass density of 0.01 g/cm$^3$. These values were obtained using the doubly-recursive method described in Eq. (\ref{double}) \cite{Gilleron2004}. The supershell under consideration comprises the orbitals (4p4d4f5s5p5d), and the chemical potential is set at $\mu = -895.58476$ eV. Corresponding one-electron energies and $D$-matrix elements are listed in Table \ref{tab1}. Figure \ref{fig1_bis} presents a zoomed-in view of Fig. \ref{fig1_full}, focusing on a narrower range of electron numbers. Figure \ref{fig2_real} depicts the same real part of the partition function ratio as shown in Figs. \ref{fig1_full} and \ref{fig1_bis}, but on a logarithmic scale, which is the conventional representation. The numerical results are compared with exact values for validation, and the computed data are superimposed on exact results. Notably, the ratio turns negative beyond 40 electrons. Figure \ref{fig2_real_bis} shows the absolute value of the real part of this ratio for electron counts greater than or equal to 40. Finally, Fig. \ref{fig2_imag} displays the imaginary part of the ratio $\mathcal{Z}_Q^{ab}/\mathcal{Z}_{Q-1}^{ab}$ under the same plasma conditions and using the same computation method. The numerical results are again overlaid with the exact values.

The coupled Eqs. (\ref{rea}) and (\ref{ima}) can be put in the matrix form
\begin{equation}
    \left(\begin{array}{c}
    \mathcal{A}_{Q,N}\\
    \mathcal{B}_{Q,N}\\
    \end{array}\right)=\sum_{p=0}^{\min(Q,g_N)}X_N^p\left[\begin{array}{cc}
    \cos(p\alpha_N) & -\sin(p\alpha_N)\\
    \sin(p\alpha_N) & \cos(p\alpha_N)\\
    \end{array}\right]
    \left(\begin{array}{c}
    \mathcal{A}_{Q-p,N-1}\\
    \mathcal{B}_{Q-p,N-1}\\
    \end{array}\right)
\end{equation}
where
\begin{equation}
\left[\begin{array}{cc}
    \cos(p\alpha_N) & -\sin(p\alpha_N)\\
    \sin(p\alpha_N) & \cos(p\alpha_N)\\
    \end{array}\right]
\end{equation}
is a rotation matrix of angle $p\alpha_N=pD_N^{ab}\,\tau/\hbar$. One has
\begin{equation}
    \cos(p\alpha_N)=\sum_{\substack{r=0,\\2r\leq n}}(-1)^r\bin{n}{2r}\cos^{p-2r}(\alpha_N)\sin^{2r}(\alpha_N)=\sum_{k=0}^{\lfloor\frac{p}{2}\rfloor}\left[(-1)^k\sum_{j=k}^{\lfloor\frac{p}{2}\rfloor}\binom{p}{2j}\binom{j}{k}\right]\cos^{p-2k}(\alpha_N)
\end{equation}
as well as
\begin{equation}
    \sin(p\alpha_N)=\sum_{\substack{r=0,\\ 2r+1\leq p}}(-1)^r\binom{p}{2r+1}\cos^{p-2r-1}(\alpha_N)\sin^{2r+1}(\alpha_N)
\end{equation}
and $\cos(p\alpha_N)$ and $\sin(p\alpha_N)$ can be respectively expressed as Chebyshev polynomials of the first kind \cite{Rivlin1974}:
\begin{equation}
    \cos(p\alpha_N)=T_p\left[\cos(\alpha_N)\right]
\end{equation}
with
\begin{equation}
	T_{n}(x)=\sum_{k=0}^{\left\lfloor {\frac {n}{2}}\right\rfloor }{\binom {n}{2k}}(x^{2}-1)^{k}x^{n-2k}
\end{equation}
and of the second kind
\begin{equation}
    \sin(p\alpha_N)=\sin(\alpha_N)~U_{p-1}\left[\cos(\alpha_N)\right]
\end{equation}
with
\begin{equation}
	U_{n}(x)=\sum _{k=0}^{\left\lfloor {\frac {n}{2}}\right\rfloor }{\binom {n+1}{2k+1}}(x^{2}-1)^{k}x^{n-2k}=x^{n}\sum _{k=0}^{\left\lfloor {\frac {n}{2}}\right\rfloor }{\binom {n+1}{2k+1}}(1-x^{-2})^{k}=\sum _{k=0}^{\left\lfloor {\frac {n}{2}}\right\rfloor }(-1)^{k}{\binom {n-k}{k}}(2x)^{n-2k}.
\end{equation}
The first Chebyshev polynomials of the first kind are $T_{0}=1$, $T_{1}(x)=x$, $T_{2}(x)=2x^{2}-1$, $T_{3}(x)=4x^{3}-3x$, $T_{4}(x)=8x^{4}-8x^{2}+1$ and $T_{5}(x)=16x^{5}-20x^{3}+5x$. The first $U_n$ polynomials are $U_{0}=1$, $U_{1}(x)=2x$, $U_{2}(x)=4x^{2}-1$, $U_{3}(x)=8x^{3}-4x$, $U_{4}(x)=16x^{4}-12x^{2}+1$ and $U_{5}(x)=32x^{5}-32x^{3}+6x$. The recurrence relations satisfied by the Chebyshev polynomials may be of interest here for numerical reasons or in order to derive new recursion relations. One has
\begin{equation}
	T_{n+1}(x)=2xT_{n}(x)-T_{n-1}(x),\ \ \forall n\geq 1,\ {\textrm {with}}\ T_{0}=1\ {\textrm {and}}\ T_{1}(x)=x,
\end{equation}
\begin{equation}
	U_{n+1}(x)=2xU_{n}(x)-U_{n-1}(x),\ \ \forall n\geq 1\ {\textrm {with}}\ U_{0}=1\ {\textrm {and}}\ U_{1}(x)=2x,
\end{equation}
as well as relations involving bith polynomials
\begin{equation}
	T_{n}(x)=U_{n}(x)-xU_{n-1}(x),\quad T_{n+1}(x)=xT_{n}(x)-(1-x^{2})U_{n-1}(x){\text{ and }}T_{n}'(x)=nU_{n-1}(x).
\end{equation}

Back to the absorption coefficient (see Eq. (\ref{basic})), one gets, using the doubly recursive method and introducing the dependency with respect to the number of subshells $N$ in $\mathcal{K}_Q(a,b)$:
\begin{equation}\label{quarante-six}
    \mathcal{K}_Q(a,b,N)=\Re\left\{\int_0^{\infty}e^{i\left(D_0^{ab}-D_a^{ab}\right)\tau/\hbar}\tilde{X}_a^{ab}(\beta,\tau)\mathcal{Z}_{\Xi,Q-1}^{ab}\left(g^{ab},\beta,\tau\right)e^{-i\hbar\omega\tau/\hbar}d\tau\right\}
\end{equation}
with
\begin{equation}
    \mathcal{Z}_{Q-1}^{ab}(g^{ab},\beta,\tau)=\sum_{|\vec{p}|=Q-1}\prod_{i=1}^N\left[\tilde{X}_i^{ab}(\beta,\tau)\right]^{p_i}\binom{g_i-\delta_{ia}-\delta_{ib}}{p_i}
\end{equation}
and since
\begin{equation}
    \mathcal{Z}_{Q-1}^{ab}(g^{ab},\beta,\tau)=\sum_{p_N=0}^{g_N}\binom{g_N-\delta_{Na}-\delta_{Nb}}{p_N}\left[\tilde{X}_N^{ab}(\beta,\tau)\right]^{p_N}\sum_{|\vec{p}|=Q-1-p_N}\prod_{i=1,\\i\ne N}^N\left[\tilde{X}_i^{ab}(\beta,\tau)\right]^{p_i}
\end{equation}
we have finally
\begin{equation}\label{quarante-neuf}
    \mathcal{K}_Q(a,b,N)=\sum_{p_N=0}^{g_N}\binom{g_N-\delta_{Na}-\delta_{Nb}}{p_N}\left[X_N^{ab}(\beta,\tau)\right]^{p_N}\mathcal{K}_Q(a,b,N-1).
\end{equation}
In practice, Eqs. (\ref{quarante-six}) to (\ref{quarante-neuf}) are efficiently computed using the techniques described above.

\begin{figure}[!ht]
 \begin{minipage}[c]{0.48\textwidth}
   \centering
   \includegraphics[width=\textwidth,angle=\anglefig,scale=\scalefig]{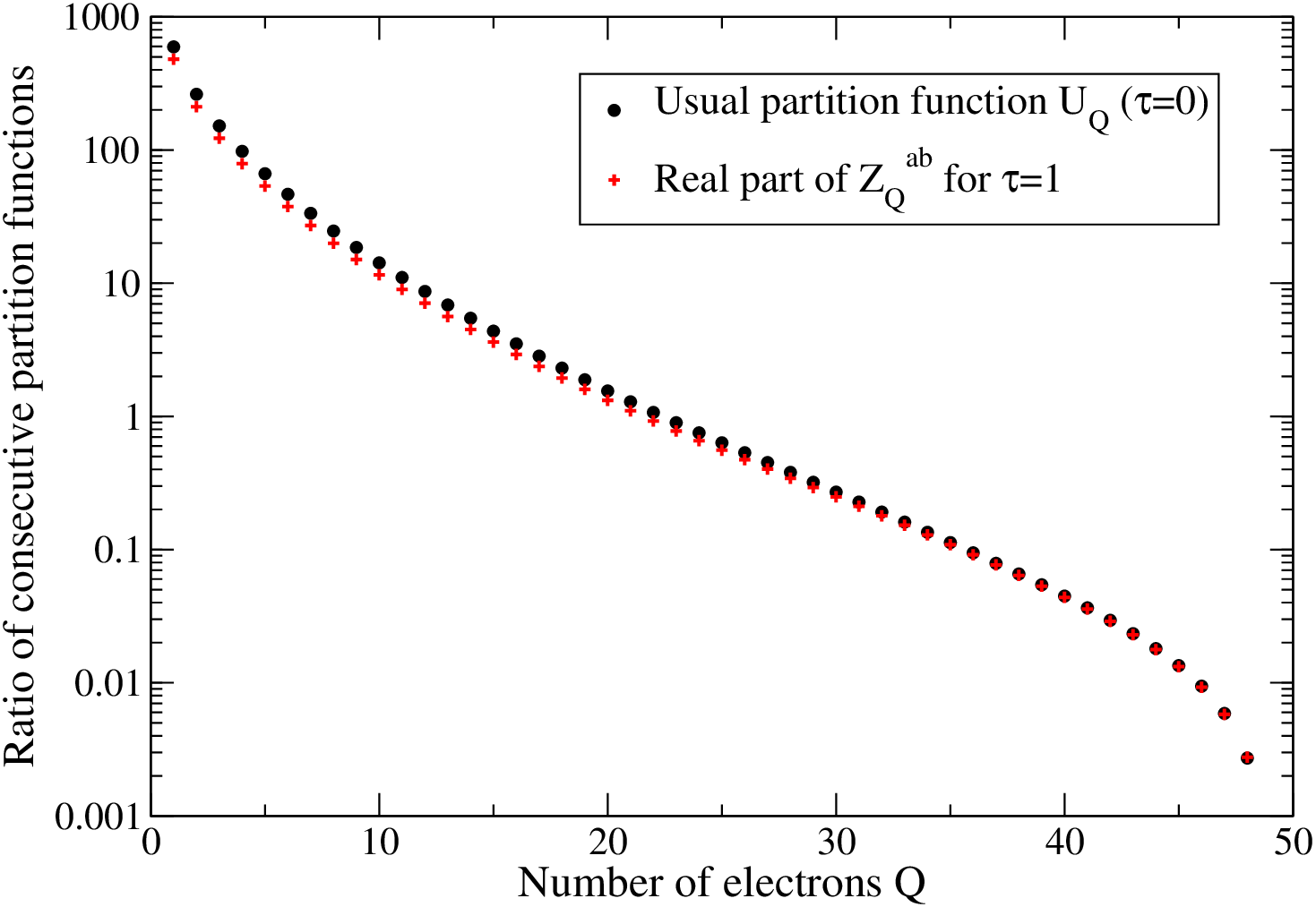}
   \caption{Comparison between ratios of consecutive partition functions obtained with the usual partition function $\mathcal{U}_Q=\mathcal{Z}_Q^{ab}(\beta,0)$ and with $\mathcal{Z}_Q^{ab}(\beta,1)$ (corresponding to $\tau$=1).}\label{fig5}
 \end{minipage}\hfill
 \begin{minipage}[c]{0.48\textwidth}
   \centering
   \includegraphics[width=\textwidth,angle=\anglefig,scale=\scalefig]{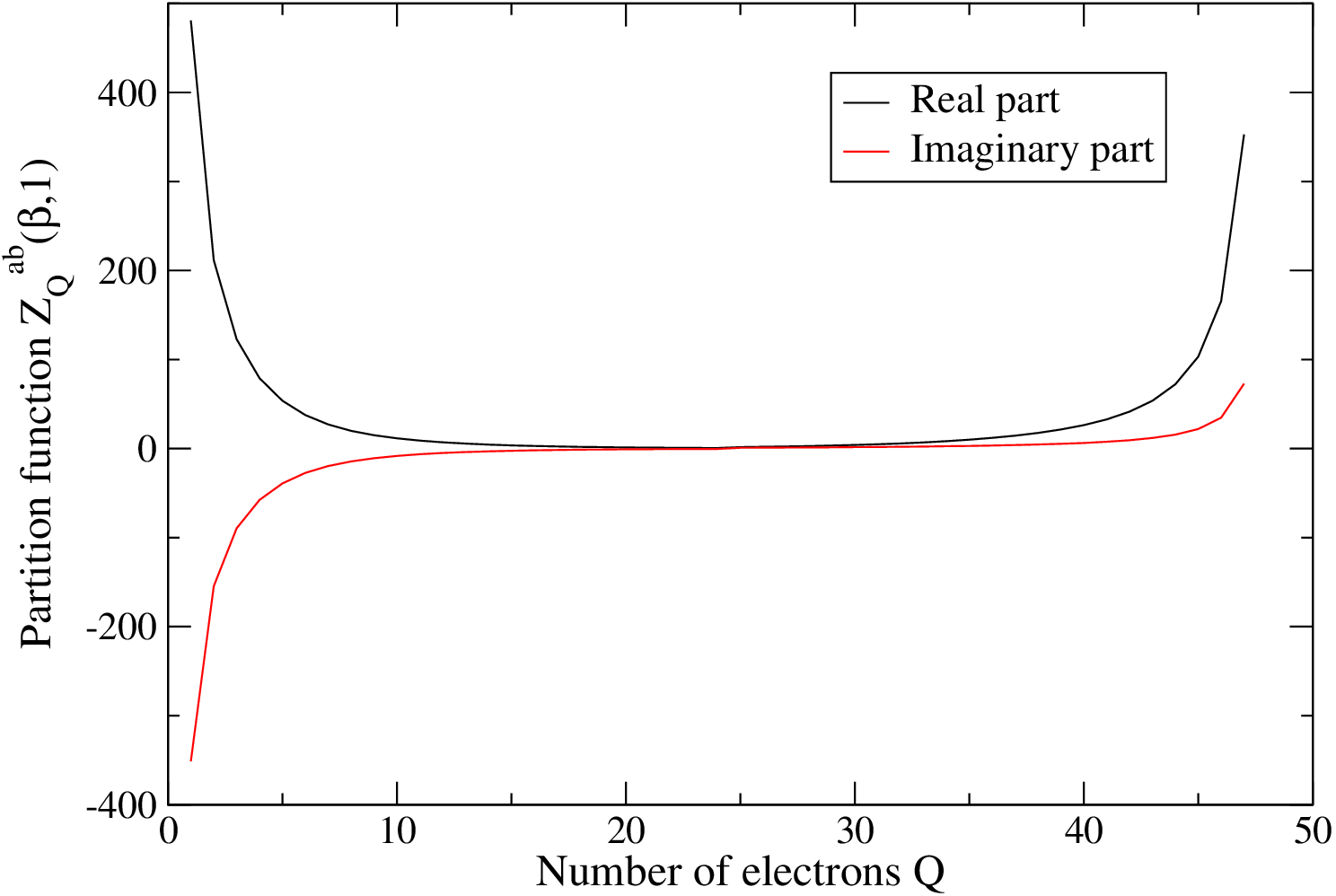}
   \caption{Real and imaginary parts of $\mathcal{Z}_Q^{ab}(\beta,1)$.}\label{fig6}
 \end{minipage}
\end{figure}

\begin{figure}[!ht]
 \begin{minipage}[c]{0.48\textwidth}
   \centering
   \includegraphics[width=\textwidth,angle=\anglefig,scale=\scalefig]{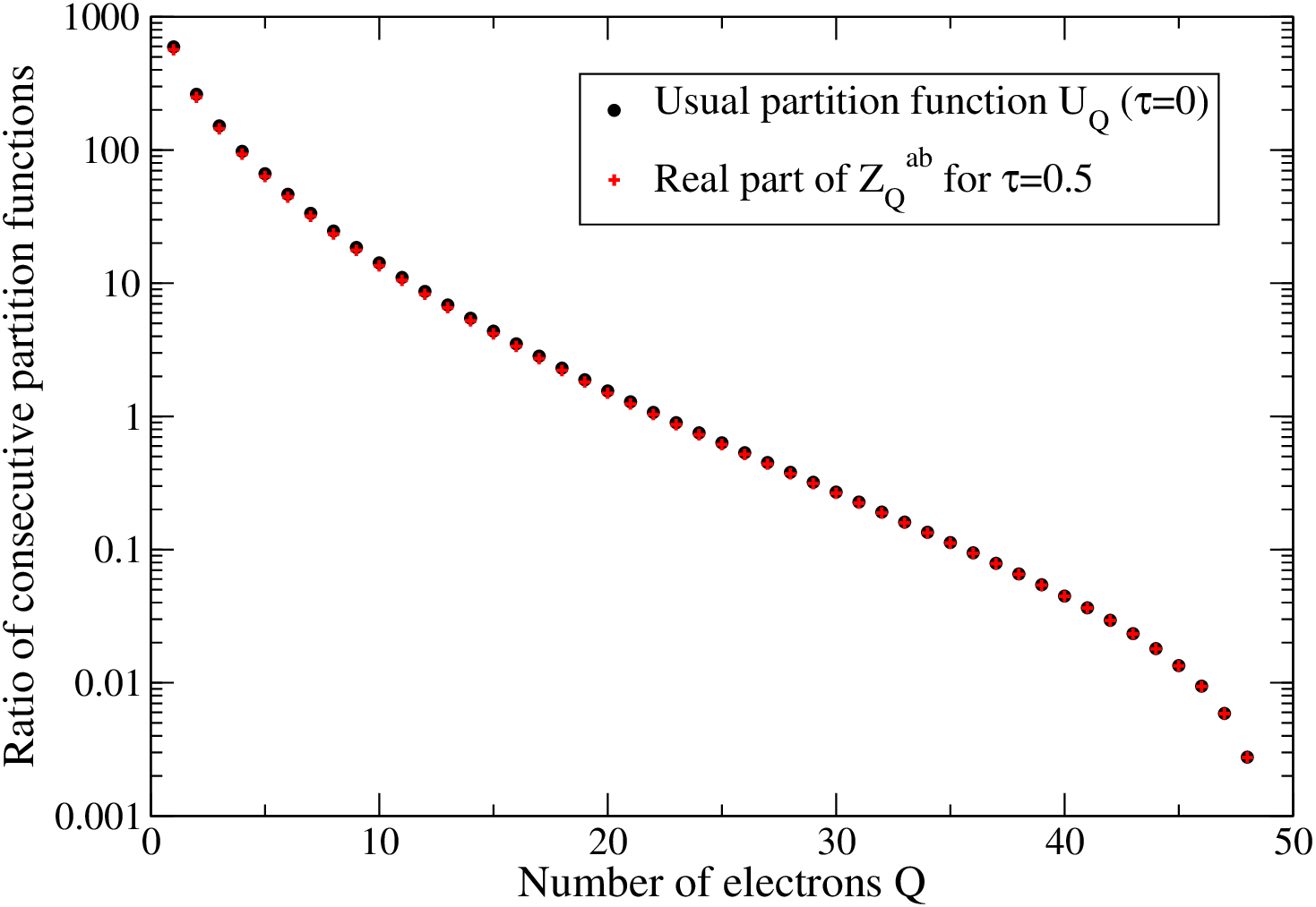}
   \caption{Comparison between ratios of consecutive partition functions obtained with the usual partition function $\mathcal{U}_Q=\mathcal{Z}_Q^{ab}(\beta,0)$ and with $\mathcal{Z}_Q^{ab}(\beta,0.5)$ (corresponding to $\tau$=0.5).}\label{fig7}
 \end{minipage}\hfill
 \begin{minipage}[c]{0.48\textwidth}
   \centering
   \includegraphics[width=\textwidth,angle=\anglefig,scale=\scalefig]{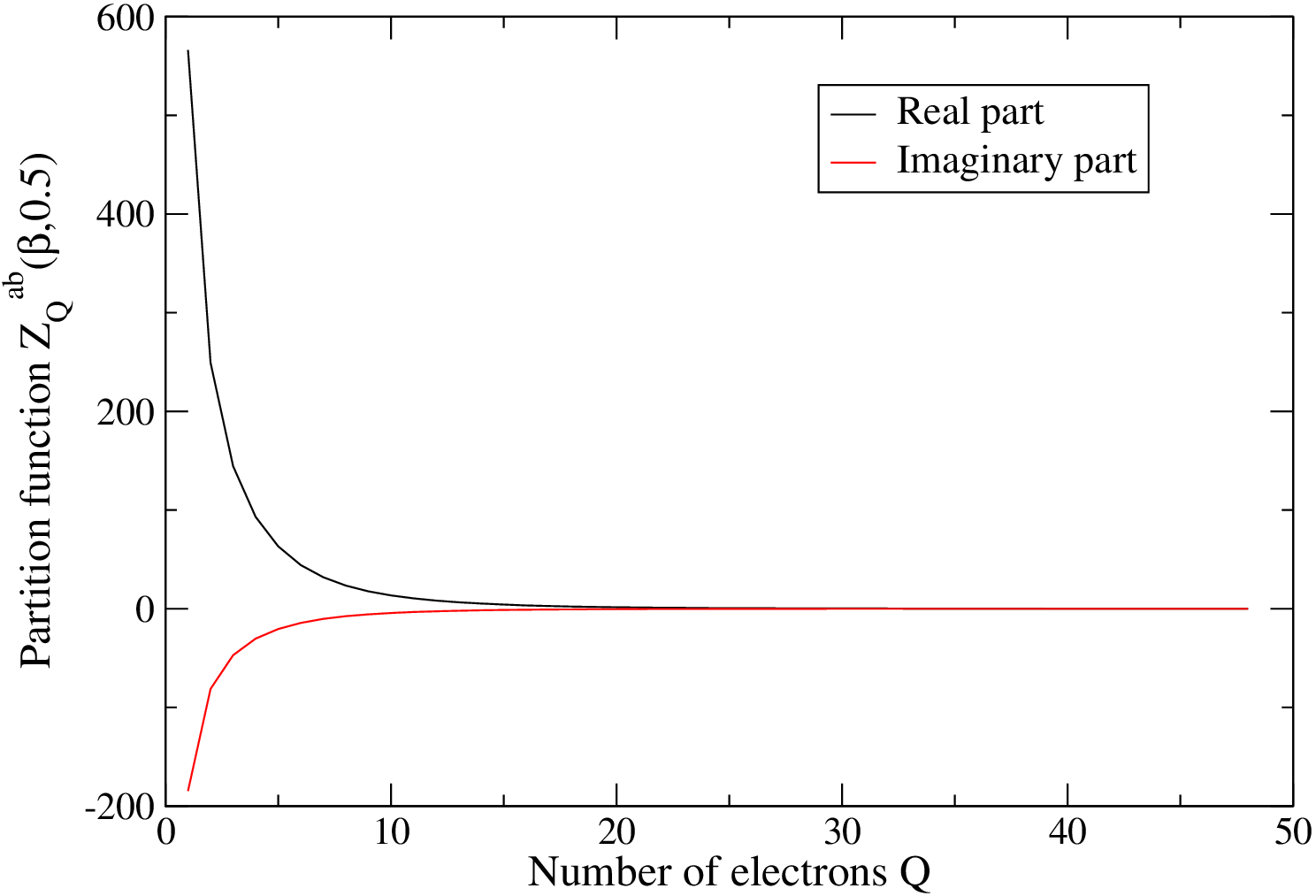}
   \caption{Real and imaginary parts of $\mathcal{Z}_Q^{ab}(\beta,0.5)$.}\label{fig8}
 \end{minipage}
\end{figure}

Figures \ref{fig5} and \ref{fig6} (for $\tau$=1), as well as \ref{fig7} and \ref{fig8} (corresponding to $\tau=0.5$) show how $\mathcal{Z}_Q^{ab}(\beta,\tau)$ tends to the usual partition function $\mathcal{U}_Q$ (independent on $a$ and $b$), when $\tau\rightarrow 0$.

\vspace{1cm}

If computation time is a major constraint, it may be relevant to resort to fast exponentiation or fast multiplication algorithms. For instance, the Sch\"onhage-Strassen algorithm is an asymptotically fast multiplication scheme for large integers \cite{Schonhage1971}. It consists in applying Fast Fourier Transform (FFT) \cite{Bailey1988} over the integers modulo $2^n+1$. The run-time bit complexity to multiply two $n$-digit numbers is $O(n.\log n.\log\log n)$. It is faster (asymptotically) than Karatsuba (in $O\left(n^{1.59}\right)$) \cite{Karatsuba1963} and Toom-Cook (in $O\left(n^{1.46}\right)$) \cite{Toom1963,Cook1966} algorithms. In 2008, F\"urer obtained an algorithm\footnote{``$\log^*$ refers to the iterated logarithm as $\log^* n=\min\left\{i\geq 0: \log^{(i)} n\leq 1\right\}$ with $\log^{(0)} n=n$ and $\log^{(i+1)} n=\log\left(\log^{(i)} n\right)$} in $O\left(n.\log n.2^{O(\log^* n)}\right)$ \cite{Furer2009}. Very recently, Harvey and van der Hoeven demonstrated that multi-digit multiplication has theoretical $O(n\log n)$ complexity \cite{Harvey2021}. However, such algorithms have been designed for the multiplication of large integers. This implies adapting our algorithm to handle integers only, incidentally performing decompositions in a particular base. However, fast-multiplication algorithms can also be applied to polynomial multiplications. They are well suited to our problem, since partition functions can be viewed as polynomials (with integer coefficients as binomial ones) in one or more variables.

\subsection{The truncated fast expansion}\label{subsec23}

In a recent work \cite{Wilson2022}, we proposed an alternative expansion for computing supershell partition functions for an arbitrary number of electrons or holes. It involves binomial coefficients and specific coefficients $\Gamma_k$. Truncating the number of terms in the expansion enables one to perform a fast approximate computation of the partition functions, which are the cornerstone of the STA method for radiative-opacity calculations \cite{Barshalom1989}. The generalization of the terminating series expansion to the complex pseudo partition functions reads
\begin{equation}\label{expan}
{\mathcal{Z}_Q}\left[ g \right] = \tilde{X}_0^Q\left\{ {\left( {\begin{array}{*{20}{c}}
G\\
Q
\end{array}} \right) + \left( {\begin{array}{*{20}{c}}
{G - 1}\\
{Q - 1}
\end{array}} \right){\tilde{\Gamma}_1} + \frac{1}{2}\left( {\begin{array}{*{20}{c}}
{G - 2}\\
{Q - 2}
\end{array}} \right){\tilde{\Gamma}_2} + \frac{1}{6}\left( {\begin{array}{*{20}{c}}
{G - 3}\\
{Q - 3}
\end{array}} \right){\tilde{\Gamma}_3} + \frac{1}{24}\left( {\begin{array}{*{20}{c}}
{G - 4}\\
{Q - 4}
\end{array}} \right){\tilde{\Gamma}_4} ...} \right\}.
\end{equation}
The $\Gamma_k$ coefficients satisfy the recurrence relation \cite{Wilson2022}:
\begin{equation}\label{recur}
    \left\{ \frac{\tilde{\Gamma}_k}{k!} \right\} = \frac{1}{k}\sum\limits_{p = 1}^k \left(-1\right)^{p+1}\tilde{\Omega}_p \,\left\{ \frac{\tilde{\Gamma}_{k - p}}{\left( k - p \right)!} \right\},
\end{equation}
with $\Gamma_0 = 1$ and
\begin{equation*}
    \tilde{\Omega}_p=\left[ {\sum\limits_{i = 1}^m {{g_i}\tilde{\Delta}_i^p} } \right],
\end{equation*}
with
\begin{equation}
\tilde{\Delta}_i=\frac{\tilde{X}_i^{ab}(\beta,\tau)}{\tilde{X}_0}-1
\end{equation}
where
\begin{equation}
\tilde{X}_0=\frac{1}{G}\sum_{i=1}^Ng_i\tilde{X}_i^{ab}(\beta,\tau).
\end{equation}

\begin{figure}[!ht]
 \begin{minipage}[c]{0.48\textwidth}
   \centering
   \includegraphics[width=\textwidth,angle=\anglefig,scale=\scalefig]{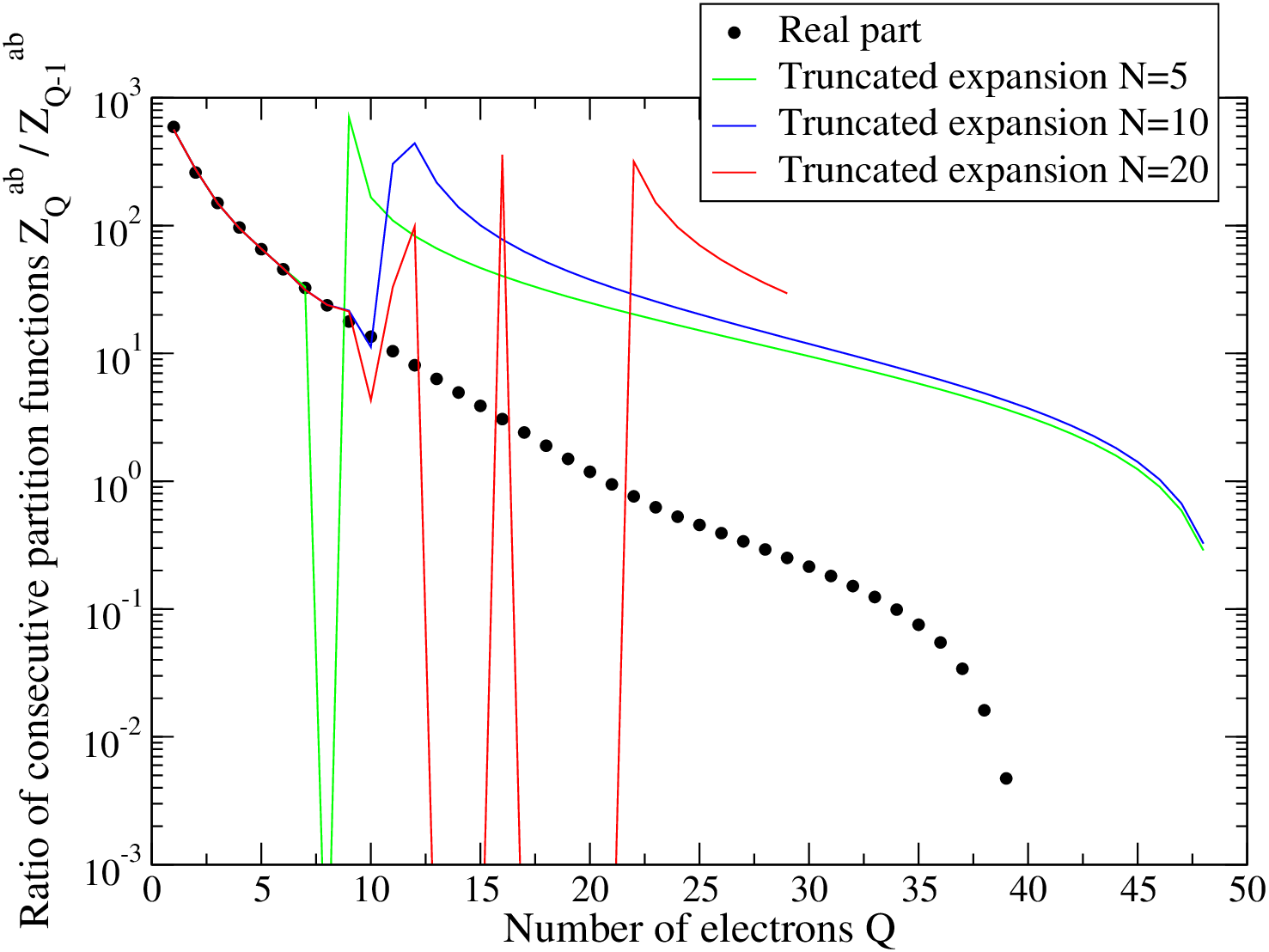}
   \caption{Real part of the ratio of consecutive partition functions $\mathcal{Z}_Q^{ab}/\mathcal{Z}_{Q-1}^{ab}$ calculated using expansion (\ref{expan}) at different orders, as a function of the number of electrons $Q$.}\label{fig4_wp_real}
 \end{minipage}\hfill
 \begin{minipage}[c]{0.48\textwidth}
   \centering
   \includegraphics[width=\textwidth,angle=\anglefig,scale=\scalefig]{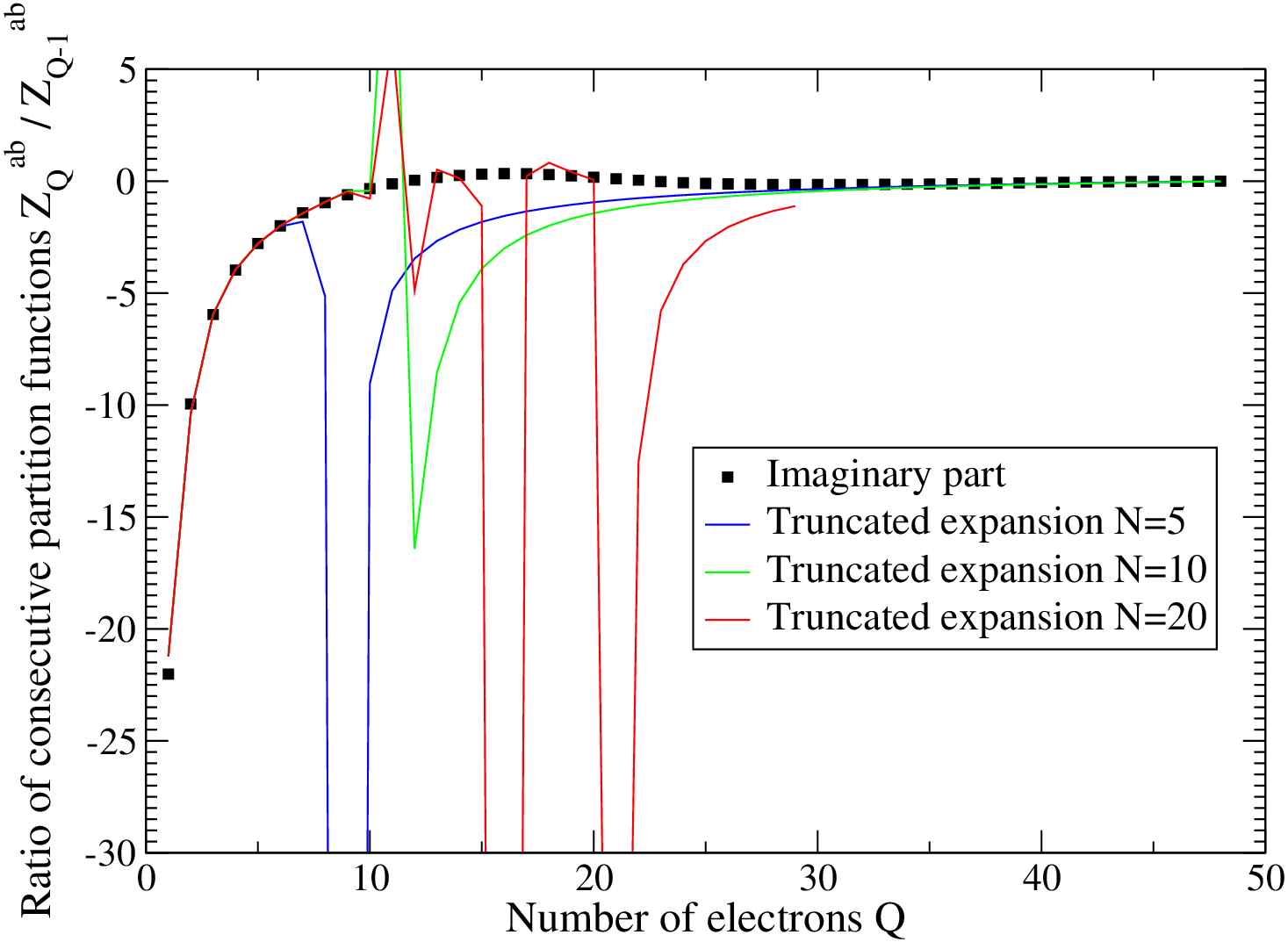}
   \caption{Imaginary part of the ratio of consecutive partition functions $\mathcal{Z}_Q^{ab}/\mathcal{Z}_{Q-1}^{ab}$ calculated using expansion (\ref{expan}) at different orders, as a function of the number of electrons $Q$.}\label{fig4_wp_imag}
 \end{minipage}
\end{figure}

An explicit formula for the $\Gamma_k$ coefficients is \cite{Pain2023}: 
\begin{equation}\label{newexpress}
    \tilde{\Gamma}_k=k!(-1)^k\sum_{\substack{\vec{q}/\\q_1+2q_2+\cdots+kq_k=k}}\prod_{p=1}^k\frac{1}{q_p!}\left(-\frac{1}{p}\sum_{i=1}^mg_i\tilde{\Delta}_i^p\right)^{q_p}.
\end{equation}
As we can see in Figs. \ref{fig4_wp_real} and \ref{fig4_wp_imag}, the brute-force implementation of the expansion yields to numerical instabilities, even with 20 terms. We also tried using quadruple precision for the complex numbers, but the numerical problems remain. As in the exact case (see section \ref{subsec22}), it is possible to separate the real and imaginary parts in the truncated expansion. One first has 
\begin{eqnarray}
	\tilde{\Delta}_j^p&=&\left[\frac{X_j}{X_0}e^{i(\alpha_j-\alpha_0)}-1\right]^p\nonumber\\
&=&\sum_{r=0}^p(-1)^{p-r}\bin{p}{r}\left\{\cos\left[r(\alpha_j-\alpha_0)\right]+i\,\sin\left[r(\alpha_j-\alpha_0)\right]\right\}
\end{eqnarray}
and thus $\sum_{j=1}^Ng_j\tilde{\Delta}_j^p=z_p+i\,\lambda_p$, with
\begin{equation}\label{zp}
	z_p=\sum_{j=1}^N\sum_{r=0}^pg_j(-1)^{p-r}\bin{p}{r}\left(\frac{X_j}{X_0}\right)^ r\cos\left[r(\alpha_j-\alpha_0)\right]
\end{equation}
and
\begin{equation}\label{lambdap}
	\lambda_p=\sum_{j=1}^N\sum_{r=0}^pg_j(-1)^{p-r}\bin{p}{r}\left(\frac{X_j}{X_0}\right)^ r\sin\left[r(\alpha_j-\alpha_0)\right].
\end{equation}
Setting
\begin{equation}	
	\frac{1}{k!}\left(V_k+i\,W_k\right)=\frac{1}{k}\sum_{p=1}^k(-1)^{p+1}\left(z_p+i\,\lambda_p\right)\frac{1}{(k-p)!}\left(V_{k-p}+i\,W_{k-p}\right),
\end{equation}
one can write, separating the real and imaginary parts:
\begin{equation}\label{vk}
	\frac{1}{k!}V_k=\frac{1}{k}\sum_{p=1}^k\frac{(-1)^{p+1}}{(k-p)!}\left(z_pV_{k-p}-\lambda_p W_{k-p}\right)
\end{equation}
as well as
\begin{equation}\label{wk}
	\frac{1}{k!}W_k=\frac{1}{k}\sum_{p=1}^k\frac{(-1)^{p+1}}{(k-p)!}\left(z_pW_{k-p}+\lambda_p V_{k-p}\right).
\end{equation}
Equations (\ref{vk}) and (\ref{wk}) can be recast in a matrix form
\begin{equation}\label{cross}
	\frac{1}{k!}\left(
	\begin{array}{c}
	V_k\\
	W_k
	\end{array}
	\right)=\frac{1}{k}\sum_{p=1}^k(-1)^{p+1}\left(
	\begin{array}{cc}
	z_p & -\lambda_p\\
	\lambda_p & z_p
	\end{array}
	\right)
	\left(
	\begin{array}{c}
	V_{k-p}\\
	W_{k-p}
	\end{array}
	\right).
\end{equation}

Examination of expressions (\ref{zp}) and (\ref{lambdap}) reveals that matrix involved in Eq. (\ref{cross}) is not a rotation matrix anymore. However, this separation between real and imaginary parts should solve the issue of numerical instability.

\section{Case of the Partially-Resolved CRSTA method}\label{sec3}

The Partially Resolved Transition Array (PRTA) approach was introduced by Iglesias and Sonnad \cite{Iglesias2012}. The main idea is to replace a transition array by a reduced one, containing only the active subshells and the subshells strongly coupled to them, with their respective populations. The remaining populated subshells are then included in the calculation as a convolution width, based on a dressing function assumed to be Gaussian (as in the original UTA formalism). A few years later, such an approach was generalized to the STA approach \cite{Wilson2015,Pain2015}, by discarding an entire supershell (at least in Ref. \cite{Pain2015}, the formalism of Ref. \cite{Wilson2015} proceeds differently). The latter variant was called ``Super-PRTA'' formalism. The combination of the PRTA to the CRSTA methods (PRTA-CRSTA) \cite{Kurzweil2016} is an improvement of the Super-PRTA one, in which the Gaussian dressing function is replaced by a CRSTA. Kurzweil and Hazak showed how to calculate, efficiently, the summation
\begin{equation}
    {\displaystyle\sum\limits_{\bar{C}_{ab}}}\underset{s\neq a,b}{\prod}\binom{g_{s}}{q_{s}}\left[\widetilde{X}_{s}\right]^{q_{s}}f_{s}\left(g_{s},q_{s},\tau\right),
\end{equation}
where
\begin{equation}
    f_{s}\left(g_{s},q_{s},\tau\right)=\exp\left[-\frac{1}{2\hbar^{2}}\left(q_{s}-\delta_{sa}\right)\left(g_{s}-q_{s}-\delta_{sb}\right)\left(\Delta_{s}^{ab}\right)^{2}\tau\right],
\end{equation}
with
\begin{equation}
\bar{C}_{ab}=\prod_{s\ne a,b}(n_s,\ell_s,j_s)^{q_s}
\end{equation}
is a configuration which does not contain the active subshells. and $\Delta_{s}^{ab}$ is given in Appendix A of Ref. \cite{Barshalom1995}. 

Figure \ref{example1_crsta} represents the transition array Fe X, [Ne] 3s$^2$ 3p$^3$ 3d$^1$ (spect)$^1$ $\rightarrow$ [Ne] 3s$^2$ 3p$^2$ 3d$^2$ (spect)$^1$, where (spect) = (4s, 4p, ... , 6p), calculated using different approaches: pure DLA (Detailed Line Accounting), which can be considered as the ``exact'' calculation, STA/SOSA with Spin-Orbit Split Arrays (SOSA) \cite{Bauche1982} statistical modeling of the transition arrays covered by the STA, STA/UTA and PRTA-CRSTA. We can see that the latter approach provides a much better description than the above mentioned statistical methods. The temperature is taken to be $T$=50 eV. The corresponding reduced transition array (i.e., without spectators), is represented in Fig. \ref{example1_crsta_nospect}. 

Figure \ref{example2_crsta} addresses the case of two electrons in the passive supershell, for transition array Fe X, [Ne] 3s$^2$ 3p$^1$ 3d$^1$ (spect)$^2$ $\rightarrow$ [Ne] 3s$^2$ 3p$^0$ 3d$^2$ (spect)$^2$. The corresponding reduced transition array (i.e., without spectators), is represented in Fig. \ref{example2_crsta_nospec}. 

In the present work, we only considered the modeling of lines or groups of lines, and did not mention the issue of spectral line shapes. Usually, the later are modeled by a Voigt function, which is the convolution of a Gaussian (for Doppler effect) and a Lorentzian (for natural width and electron collisions), yielding
\begin{equation}
H(a,v)=\frac{a}{\pi}\int_{-\infty}^ {\infty}\frac{e^{-y^2}}{(v-y)^2+a^2}dy=\frac{1}{\sqrt{\pi}}\int_0^{\infty}\exp\left(-ax-\frac{x^ 2}{4}\right)\cos(vx)dx,
\end{equation}
where $a$ is the ratio of the Lorentzian Full Width at Half Maximum (FWHM) to the Gaussian FWHM, and $v$ the distance from the line center in units of the Doppler FWHM. It is well known that the Voigt function has a nonphysical asymptotic behavior. In the case where the separation between lines is large enough, this nonphysical tail can lead to significant error in the evaluated opacity. A solution to this problem may be found in the use of truncation methods, applied in the frequency domain \cite{Iglesias2009}. As pointed out by Kurzweil and Hazak \cite{Kurzweil2016}, the present treatment in the time domain opens the way to another convenient approach which does not suffer from the nonphysical large-tail problem, by the application of collision-narrowed line profiles, which in the time domain have the form
\begin{equation}
    \exp\left\{-\frac{\gamma\tau}{\hbar}-\frac{\alpha^2}{\hbar^2\eta^2}\left[\eta\tau-1+\exp(-\eta\tau)\right]\right\}, 
\end{equation}
where $\eta$ is the coefficient of the dynamical friction undergone by the moving emitting particle and $(\alpha,\gamma)$ are parameters given in Ref. \cite{Galatry1961}. 

\begin{figure}[!ht]
 \begin{minipage}[c]{0.48\textwidth}
   \centering
   \includegraphics[width=\textwidth,angle=\anglefig,scale=\scalefig]{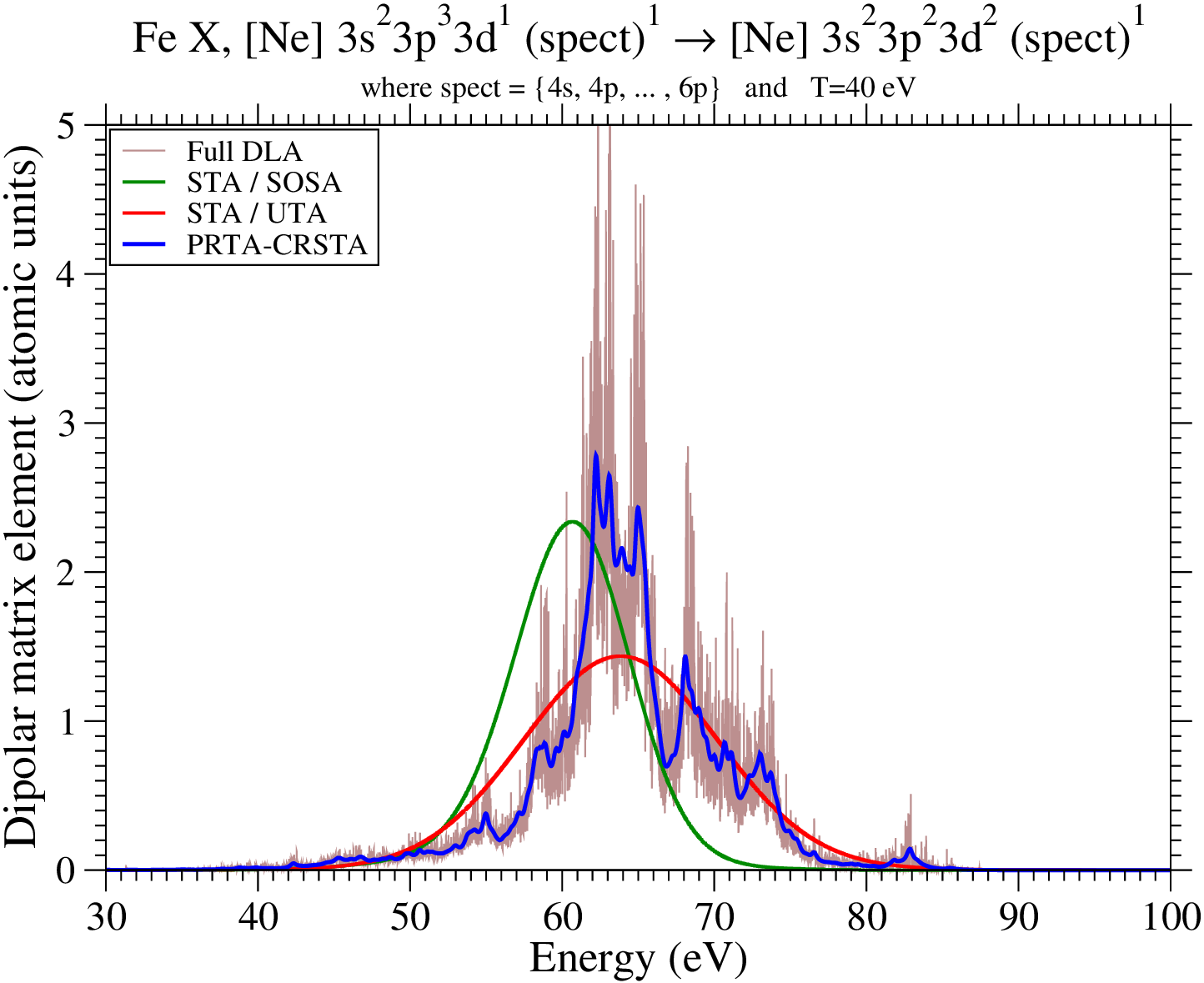}
   \caption{Transition array Fe X, [Ne] 3s$^2$ 3p$^3$ 3d$^1$ (spect)$^1$ $\rightarrow$ [Ne] 3s$^2$ 3p$^2$ 3d$^2$ (spect)$^1$, where (spect) = (4s, 4p, ... , 6p), calculated using different approaches: pure DLA (Detailed Line Accounting), STA/SOSA, STA/UTA and PRTA-CRSTA. The temperature of the iron plasma is taken to be $T$=50 eV.}\label{example1_crsta}
 \end{minipage}\hfill
 \begin{minipage}[c]{0.48\textwidth}
   \centering
   \includegraphics[width=\textwidth,angle=\anglefig,scale=\scalefig]{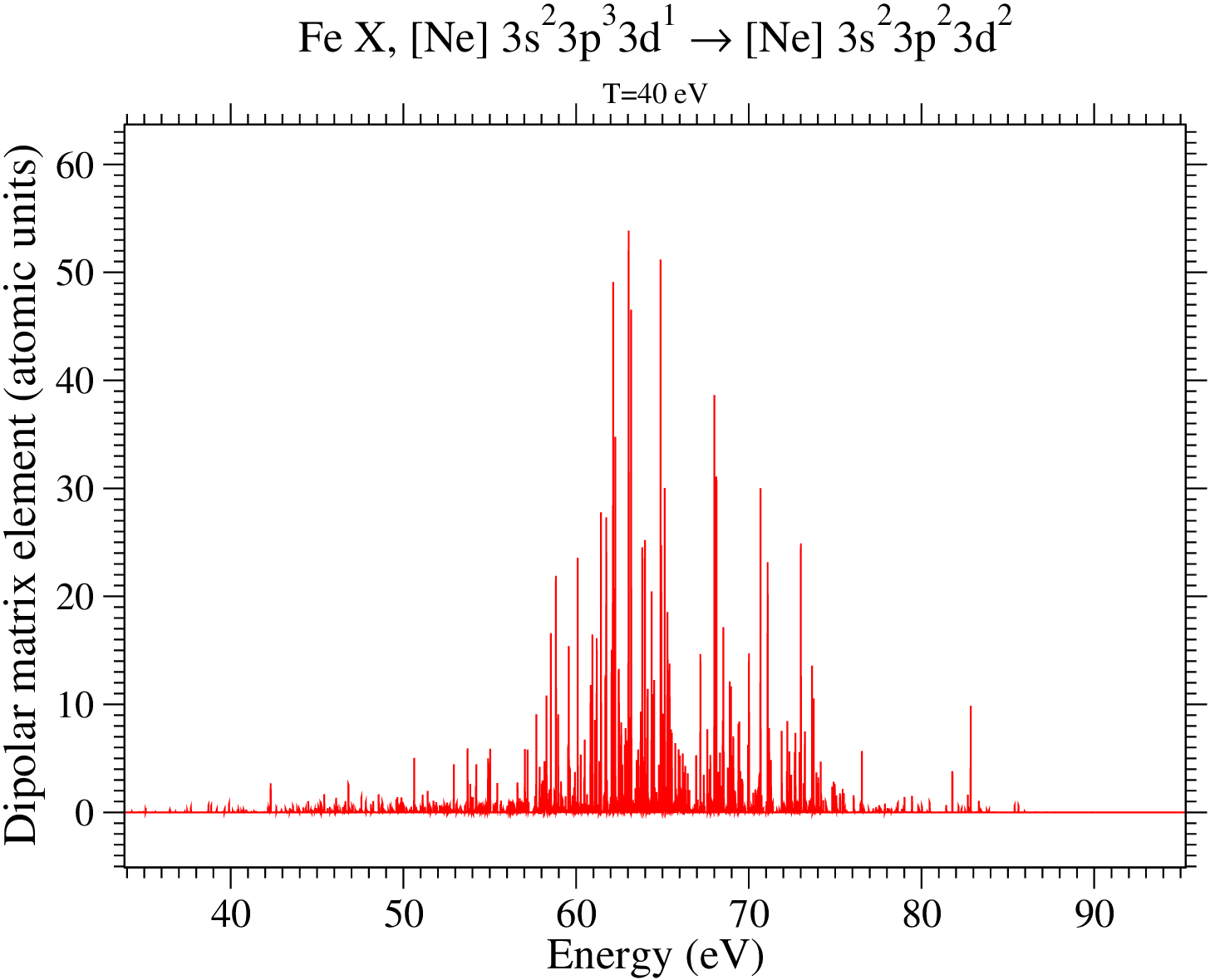}
   \caption{Reduced transition array (i.e., without spectators) corresponding to the case of Fig. \ref{example1_crsta_nospect}.}\label{example1_crsta_nospect}
 \end{minipage}
\end{figure}

\begin{figure}[!ht]
 \begin{minipage}[c]{0.48\textwidth}
   \centering
   \includegraphics[width=\textwidth,angle=\anglefig,scale=\scalefig]{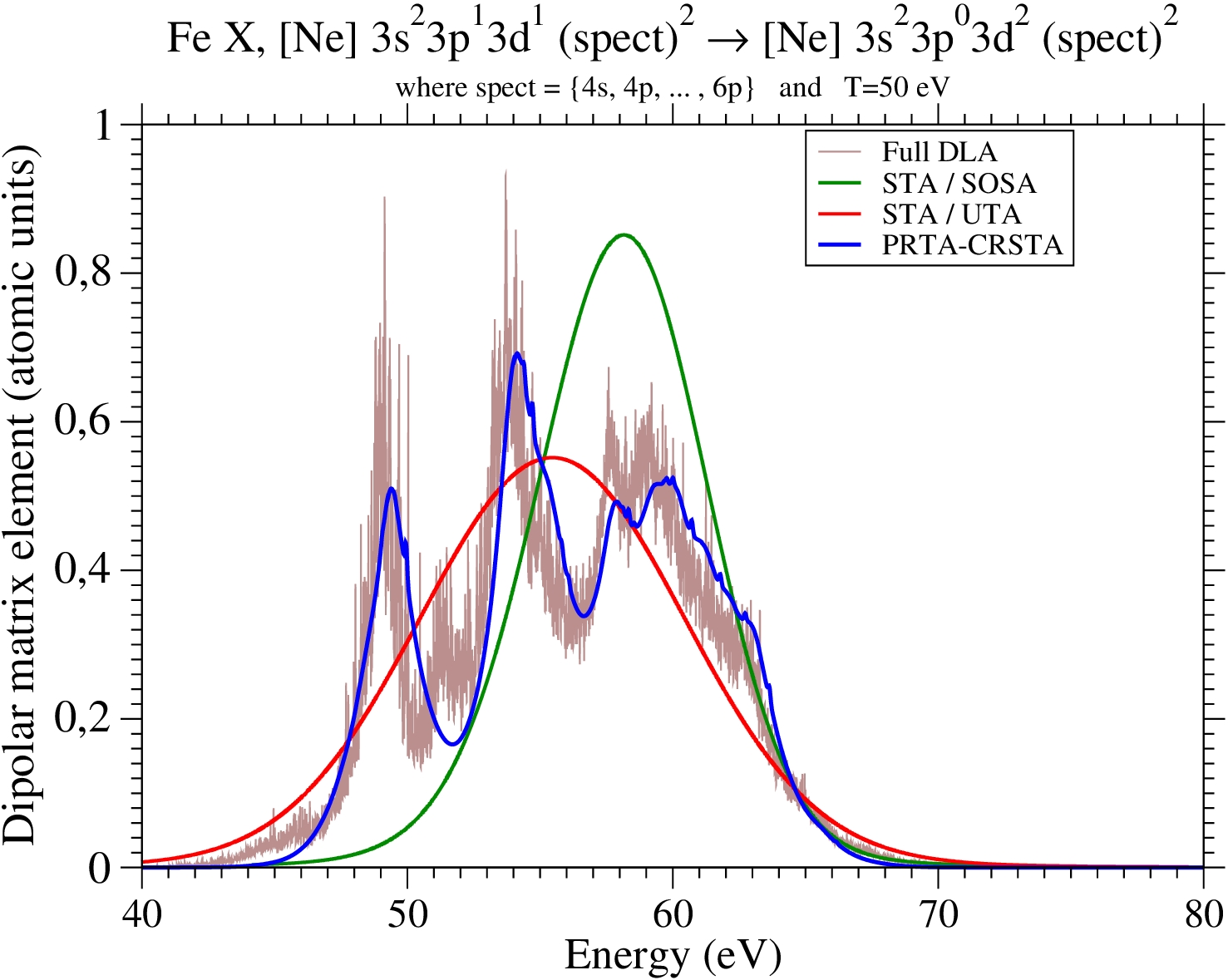}
   \caption{Transition array Fe X, [Ne] 3s$^2$ 3p$^1$ 3d$^1$ (spect)$^2$ $\rightarrow$ [Ne] 3s$^2$ 3p$^0$ 3d$^2$ (spect)$^2$, where (spect) = (4s, 4p, ... , 6p), calculated using different approaches: pure DLA (Detailed Line Accounting), STA/SOSA, STA/UTA \cite{Bauche1982} statistical modeling of the transition arrays covered by the STA, STA/UTA and PRTA-CRSTA. The temperature of the iron plasma is taken to be $T$=50 eV.}\label{example2_crsta}
 \end{minipage}\hfill
 \begin{minipage}[c]{0.48\textwidth}
   \centering
   \includegraphics[width=\textwidth,angle=\anglefig,scale=\scalefig]{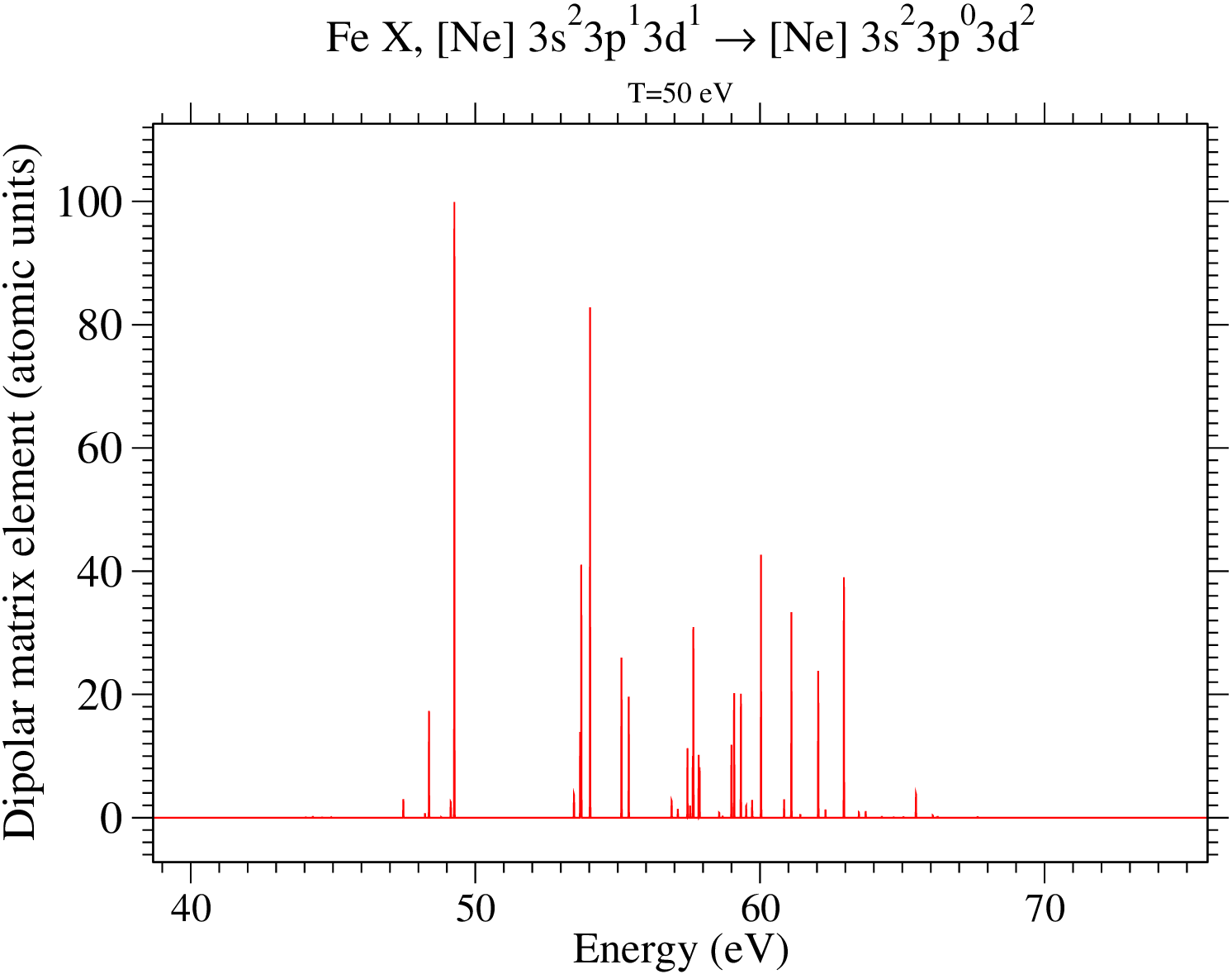}
   \caption{Reduced transition array (i.e., without spectators) corresponding to the case of Fig. \ref{example2_crsta}.}\label{example2_crsta_nospec}
 \end{minipage}
\end{figure}

\section{Conclusion}

In the usual Super-Transition-Arrays approach for radiative opacity, all the transition arrays associated to a given one-electron jump are represented by a Gaussian function, which intensity, mean energy and variance can be calculated analytically using a partition-function algebra. Kurzweil and Hazak developed a new method, called ``Configurationally Resolved Super-Transition-Arrays'' (CRSTA), for the calculation of the spectral opacity of hot plasmas. In the latter approach, the spectrum of each super transition array is computed as the Fourier transform of a single complex pseudo-partition function, which represents the exact analytical summation of the contributions of all constituting unresolved transition arrays sharing the same set of one-electron states. Therefore, in such a new method, the spectrum of each super transition array is resolved down to the level of the (unresolved) transition arrays. It is shown that the corresponding spectrum, computed by the traditional STA method, is in fact just the coarse-grained Gaussian approximation of the CRSTA. The authors developed a new computer program, able to evaluate the absorption coefficient by both the new configurationally resolved and the traditional Gaussian STA model. In this article, we found that, despite the imaginary part, the doubly-recursive relation - over both the numbers of electrons and subshells -, which was introduced to avoid problems due to alternating-sign terms in the standard STA formalism, remains numerically stable. This was rather unexpected, in particular because of the occurrence of trigonometric functions, leading to sign changes. The recursion relation boils down to two other ones, coupling the real and imaginary parts of the partition functions. This is beneficial for several reasons: we can deal only with recursion relations on real numbers (which is more practical, in particular in terms of numerical accuracy), the formalism lends itself well to matrix treatment and the calculations can be made easier using properties of Chebyshev polynomials. The numerical stability is mostly due to the fact that, in the complex case, the recursion relation can be presented in a form involving rotation matrices. The vector of real and imaginary parts at a given iteration is therefore obtained as a sum of the rotated previous ones. On the contrary, our recently published approximate method, relying on a truncated expansion and on the theory of partitions, can not be applied as it to the CRSTA method. It was found to be subject to numerical instabilities preventing convergence with respect to the number of terms in the expansion, even using quadruple precision. 




\appendix

\section{Hartree-Fock energy of a configuration}\label{appA}

The Hartree-Fock energy of a configuration reads
\begin{equation}
    E_c=\sum_{s=1}^Nq_s\left(\frac{1}{2}(q_s-1)V_{s,s}+\frac{1}{2}\sum_{\substack{r=1,\\ r\ne s}}^Nq_rV_{s,r}\right)
\end{equation}
or equivalently (using the fact that $G^{(k)}(n_s\ell_s,n_s\ell_s)=F^{(k)}(n_s\ell_s,n_s\ell_s)$):
\begin{eqnarray}
    E_c&=&\frac{1}{2}\sum_{s=1}^Nq_s(q_s-1)\left[F^{(0)}(n_s\ell_s,n_s\ell_s)-\frac{g_s}{2(g_s-1)}\sum_{k>0}\threejm{\ell_s}{k}{\ell_s}{0}{0}{0}^2G^{(k)}(n_s\ell_s,n_s\ell_s)\right]\nonumber\\
    & &+\frac{1}{2}\sum_{s=1}^Nq_sq_r\left[F^{(0)}(n_s\ell_s,n_r\ell_r)-\frac{1}{2}\sum_{k=0}\threejm{\ell_s}{k}{\ell_r}{0}{0}{0}^2G^{(k)}(n_s\ell_s,n_r\ell_r)\right].
\end{eqnarray}
It can be recast in the form
\begin{eqnarray}
    E_c&=&\frac{1}{2}\sum_{s=1}^Nq_s(q_r-\delta_{s,r})\left[F^{(0)}(n_s\ell_s,n_r\ell_r)-\frac{g_s}{2(g_s-\delta_{s,r})}\sum_{k>0}\threejm{\ell_s}{k}{\ell_r}{0}{0}{0}^2G^{(k)}(n_s\ell_s,n_r\ell_r)\right]\nonumber\\
    & &-\frac{1}{2}\sum_{\substack{s,r=1,\\r\ne s}}^Nq_sq_r\frac{1}{2}\delta_{\ell_s,\ell_r}\frac{1}{2\ell_s+1}G^{(0)}(n_s\ell_s,n_r\ell_r)
\end{eqnarray}
yielding
\begin{eqnarray}
    E_c&=&\frac{1}{2}\sum_{s,r=1}^Nq_s(q_r-\delta_{r,s})\left[F^{(0)}(n_s\ell_s,n_r\ell_r)-(1-\delta_{s,r})\frac{\delta_{\ell_s,\ell_r}}{g_s}G^{(0)}(n_s\ell_s,n_r\ell_r)\right.\nonumber\\
    & &\left.-\frac{g_s}{2(g_s-\delta_{s,r})}\sum_{k>0}\threejm{\ell_s}{k}{\ell_r}{0}{0}{0}^2G^{(k)}(n_s\ell_s,n_r\ell_r)\right].
\end{eqnarray}
Therefore, one can write
\begin{equation}
    E_c=\frac{1}{2}\sum_{r,s=1}^Nq_s(q_r-\delta_{s,r})\tilde{V}_{s,r}
\end{equation}
with
\begin{equation}
    \tilde{V}_{s,r}=F^{(0)}\left(n_s\ell_s,n_r\ell_r\right)-(1-\delta_{s,r})\frac{\delta_{\ell_r,\ell_s}}{g_s}G^{(0)}(n_s\ell_s,n_r\ell_r)-\frac{g_s}{2(g_s-\delta_{s,r})}\sum_{k>0}\threejm{\ell_s}{k}{\ell_r}{0}{0}{0}^2G^{(k)}(n_s\ell_s,n_r,\ell_r).
\end{equation}

\section{Proof of the Bar-Shalom relation}\label{appB}

The recurrence relation proposed by Bar-Shalom is explained in the Appendix A of the 1989 article. However, it contains many shortcuts. We provide here a more detailed proof, for pedagogical purpose. The generating function associated to the counting problem is
\begin{equation}
    F(z)=\sum_Qz^Q\mathcal{U}_Q(g)=\prod_{s=1}^N\left(1+zX_s\right)^{g_s}
\end{equation}
and 
\begin{equation}
    \mathcal{U}_Q=\left.\frac{1}{Q!}\frac{d^Q}{dz^Q}F(z)\right|_{z=0}
\end{equation}
which can be put in the form
\begin{equation}
    \mathcal{U}_Q=\left.\frac{1}{Q}\left[\frac{1}{(Q-1)!}\frac{d^{Q-1}}{dz^{Q-1}}\sum_{r=1}^Ng_rX_r\prod_{s=1}^N\left(1+zX_s\right)^{g_s-\delta_{s,r}}\right]\right|_{z=0}.
\end{equation}
The latter identity can be recast into
\begin{equation}
    \mathcal{U}_Q=\left.\frac{1}{Q}\left[\frac{1}{(Q-1)!}\frac{d^{Q-1}}{dz^{Q-1}}\left(\sum_{r=1}^Ng_rX_r\prod_{s=1}^N\left(1+zX_s\right)^{g_s}-z\sum_{r=1}^Ng_rX_r^2\prod_{s=1}^N\left(1+zX_s\right)^{g_s-\delta_{s,r}}\right)\right]\right|_{z=0}.
\end{equation}
Using the relation
\begin{equation}
    \left.\frac{1}{(Q-1)!}\frac{d^{Q-1}}{dz^{Q-1}}~zF(z)\right|_{z=0}=\left.\binom{Q-1}{1}\frac{d^{Q-2}}{dz^{Q-2}}~zF(z)\right|_{z=0}
\end{equation}
we get
\begin{equation}
    \mathcal{U}_Q(g)=\left.\frac{1}{Q}\left[\vphantom{\prod_{i=1}^N}\chi_1~\mathcal{U}_{Q-1}(g)-\frac{1}{(Q-2)!}\frac{d^{Q-2}}{dz^{Q-2}}\sum_{r=1}^Ng_rX_r^2\prod_{s=1}^N\left(1+zX_s\right)^{g_s-\delta_{s,r}}\right]\right|_{z=0}
\end{equation}
and by repeated application of the latter procedure, one finds
\begin{eqnarray}
    \mathcal{U}_Q(g)&=&\frac{1}{Q}\left[\chi_1~\mathcal{U}_{Q-1}(g)-\chi_2 \mathcal{U}_{Q-2}(g)\right.\nonumber\\
    & &\left.\left.+\frac{1}{(Q-3)!}\frac{d^{Q-3}}{dz^{Q-3}}\sum_{r=1}^Ng_rX_r^3\prod_{s=1}^N\left(1+zX_s\right)^{g_s-\delta_{s,r}}\right]\right|_{z=0},
\end{eqnarray}
and so on.

\end{document}